\documentclass[twocolumn, aps, pra, superscriptaddress]{revtex4-1}
\usepackage{amsfonts, amssymb, amsmath, graphicx, comment, bm, slashed, dcolumn, color}
\usepackage{wasysym}

\begin{document}

\title{Superfluid transition temperature and fluctuation theory of
\\  spin-orbit and  Rabi coupled fermions with tunable interactions}

\author{Philip D. Powell}
\affiliation{Lawrence Livermore National Laboratory, 7000 East Avenue, 
Livermore, California 94550, USA}
\affiliation{Department of Physics, University of Illinois at Urbana-Champaign, 
1110 W. Green Street, Urbana, Illinois 61801, USA}

\author{Gordon Baym}
\affiliation{Department of Physics, University of Illinois at Urbana-Champaign, 
1110 W. Green Street, Urbana, Illinois 61801, USA}

\author{C. A. R. S\'{a} de Melo}
\affiliation{School of Physics, Georgia Institute of Technology, 
837 State Street, Atlanta, Georgia 30332, USA}

\date{\today}

\begin{abstract}
We obtain the superfluid transition temperature of equal Rashba-Dresselhaus spin-orbit-
and Rabi-coupled Fermi superfluids, from the Bardeen-Cooper-Schrieffer (BCS) 
to Bose-Einstein condensate (BEC) regimes in three dimensions for tunable $s$-wave interactions. 
In the presence of Rabi coupling, we find that spin-orbit coupling enhances (reduces) the critical temperature
in the BEC (BCS) limit. For fixed interactions, we show that spin-orbit coupling can convert
a first-order (discontinuous) phase transition into a second-order (continuous) phase transition, as a function
of Rabi coupling.  We derive the Ginzburg-Landau free energy to sixth power 
in the superfluid order parameter to describe both continuous and discontinuous phase transitions
as a function of spin-orbit and Rabi couplings. Lastly, we develop a time-dependent Ginzburg-Landau fluctuation
theory for an arbitrary mixture of Rashba and Dresselhaus spin-orbit couplings at any interaction strength.
\end{abstract}

\maketitle

\section{Introduction}
The ability to simulate magnetic and other external fields~\cite{spielman-2009a,  spielman-2009b, ketterle-2015, spielman-2011, zwierlein-2012, wang-2012, spielman-2013a, galitski-2013, zhang-2014,fallani-2015, huang-2016, wu-2016}
in cold atomic gases has created the opportunity to explore a wide variety of new interactions
and complex phase structures otherwise inaccessible in the laboratory.  Moreover, the capacity
to generate these synthetic fields in both bosonic and fermionic systems, and to continuously tune
two-body interactions by means of a Feshbach resonance, has opened up a wonderland of tunable
systems, previously restricted to theorists' dreams.  For example, the possibility of simulating
quantum chromodynamics (QCD) on an optical
lattice~\cite{cirac-2012, wiese-2013, zohar-2015, dalmonte-2016} is a tantalizing prospect for
researchers whose current theoretical tools remain limited by QCD's non-perturbative character
and the restriction of lattice techniques to near-zero chemical potential.

Previous theoretical analyses of three-dimensional spin-orbit-coupled
Fermi gases (e.g., $^{6}$Li, $^{40}$K) have focused mainly on the zero-temperature limit,
in which several exotic phases characterized by unconventional pairing are expected
to emerge~\cite{zhang-2011, zhai-2011, pu-2011, han-2012a, seo-2012a, seo-2012b}.
However, the Raman laser platforms currently employed to produce synthetic
spin-orbit fields also induce heating that prevents the realization of
temperatures sufficiently low to observe
the superfluid transition in either the weakly coupled Bardeen-Cooper-Schrieffer (BCS) or
the strongly coupled Bose-Einstein condensate (BEC)
regimes~\cite{zwierlein-2012, spielman-2013a}.
Thus, while two-body bound states (Feshbach molecules) have been observed
in the BEC limit of $^{40}$K~\cite{spielman-2013a, doga-2016}, the
observation of superfluid states remain elusive. Future experiments, however,
may break this impasse by employing a new platform currently under
development---the radio-frequency atom chip---which avoids heating
of the atom cloud entirely~\cite{spielman-2010}.
While rf atom chips are somewhat more restricted than the Raman scheme
in the maximum obtainable spin-orbit coupling, its potential to reach
superfluid temperatures is leading to its adoption in the next generation
of experiments probing the topological superfluid
phases of spin-orbit-coupled fermions~\cite{spielman-private}.

One class of systems of particular interest in the context of quantum simulation
is that of Rashba-Dresselhaus spin-orbit-coupled gases~\cite{zhang-2011, zhai-2011, pu-2011, han-2012a, seo-2012a, seo-2012b, Juzeliunas-2010, campbell-2011}.
These systems are intriguing both because they reflect physics studied extensively in the 
context of semiconductors~\cite{dresselhaus-1955, rashba-1960}, 
and because they provide a platform for realizing tunable 
non-Abelian fields in the laboratory.
Thus, while the holy grail of a full optical simulation of QCD remains years in the future,
there do exist notable analogies between quark matter and cold atomic systems
(e.g., non-Abelian fields, evolution between strongly and weakly coupled limits)
within near-term experimental reach~\cite{sademelo-2008, ozawa-2010, powell-2013, doga-2018}.  
Investigations of spin-orbit-coupled ultracold gases have also included
optical lattices~\cite{yamamoto-2017, ye-2018, ye-2018a, rey-2019, yi-2019, yau-2019, liu-2019,marsiglio-2020, hofstetter-2020, hofstetter-2021}, thus enlarging the number of possible physical
systems that can be accessible experimentally.  

To date, most experimental realizations of these systems have adopted equal
Rashba-Dresselhaus couplings~\cite{spielman-2011, zwierlein-2012, wang-2012, spielman-2012},
but systems exhibiting Rashba-only couplings have also
been created~\cite{huang-2016, zhang-2016a, spielman-2021}.
Other experiments have generated spin-orbit coupling dynamically~\cite{lev-2019} or even
created three-dimensional spin-orbit coupling~\cite{pan-2021}. 
Due to the versatility of Rashba-Dresselhaus coupled 
systems, the ability to realize these systems in the laboratory, and
the myriad technical challenges
inherent in reaching arbitrarily low temperatures, it is increasingly
important to provide a theoretical framework for guiding and testing
these simulators against experimental probes at realistic (nonzero) temperatures.

This problem bears a close relation to spin-orbit coupling in solids, where the role of the
Rabi frequency is played by an external Zeeman magnetic field. 
While a mean-field treatment describes well the evolution from the
BCS to the BEC regime at zero temperature~\cite{leggett-1980, sademelo-1997},
this order of approximation fails to describe the correct critical temperature of the system 
in the BEC regime because the physics of two-body bound states, i.e., Feshbach molecules,
is not captured when the pairing order
parameter goes to zero~\cite{sademelo-1993}.
To remedy this problem, we include the effects of
order-parameter fluctuations in the thermodynamic potential.

In this paper, we investigate the impact of a specific class of spin-orbit coupling,
namely, an equal mixture of Rasha and Dresselhaus terms, on the superfluid transition
temperature of a three-dimensional Rabi-coupled Fermi gas, but also give general results
for an arbitrary mixture of Rashba and Dresselhaus components. This paper is the longer
version of our preliminary work~\cite{powell-2017}.
We stress that the present results are applicable to both neutral cold atomic and
charged condensed-matter systems.   We show that spin-orbit coupling,
in the presence of a Rabi field (or Zeeman field, in solids), enhances
the critical temperature of the superfluid in the BEC regime and
converts a discontinuous first-order phase transition into a continuous
second-order transition, as a function of the
Rabi frequency for given two-body interactions. 
We analyze the nature of the phase transition in terms of the Ginzburg-Landau
free energy, calculating it to the sixth power of the superfluid order parameter,
as required to describe both discontinuous transitions
as a function of the spin-orbit coupling, Rabi frequency, and two-body interactions.

This paper is organized as follows.
In Sec.~\ref{sec:hamiltonian-and-action}, we describe the Hamiltonian and action
for three-dimensional Fermi gases in the presence of a general Rashba-Dresselhaus
spin-orbit coupling, Rabi field, and tunable $s$-wave interactions. We also obtain the
inverse Green operator that is used in the calculation of the thermodynamic potential
and Ginzburg-Landau theory of subsequent sections.
In Sec.~\ref{sec:thermodynamic-potential}, we analyze the thermodynamic potential
across the entire BCS-to-BEC evolution, including contributions from both
the mean-field and Gaussian fluctuations, and obtain the order parameter and
number equations.
In Sec.~\ref{sec:critical-temperature},
we study the combined effects of Rabi fields and spin-orbit coupling
on the superfluid critical temperature, constructing the
finite-temperature phase diagram versus Rabi fields and scattering parameter.
In Sec.~\ref{sec:ginzburg-landau-theory}, we present the Ginzburg-Landau (GL)
theory for the superfluid order parameter and investigate further corrections
to the critical temperature in the BEC limit by including interactions between
bosonic bound states. The GL action is obtained to sixth order in the order parameter
to allow for the existence of discontinuous (first-order) phase transitions.
In Sec.~\ref{sec:comparison-to-earlier-work}, we compare our work on the
experimentally relevant equal Rashba-Dresselhaus spin-orbit coupling with
earlier work that has considered different forms of theoretically motivated
spin-orbit couplings. In Sec.~\ref{sec:conclusions}, we conclude and look
toward the future of experimental work in this field.

In the interest of readability, we relegate a number of detailed calculations to appendices.
In Appendix~\ref{sec:appendix-A}, we discuss the Hamiltonian and effective
Lagrangian for a general Rashba-Dresselhaus spin-orbit coupling.
In Appendix~\ref{sec:appendix-B}, we analyze the saddle-point
approximation for general Rashba-Dresselhaus spin-orbit coupling.
In Appendix~\ref{sec:appendix-C}, we derive the modified number equation, including
the contribution arising from Gaussian fluctuations, which renormalizes the chemical
potential obtained at the saddle-point level.
In Appendix~\ref{sec:appendix-D}, using a general Rashba-Dresselhaus
spin-orbit coupling, we obtain expressions for the coefficients of the
Ginzburg-Landau theory up to sixth order in order parameter. 

\section{Hamiltonian and Action}
\label{sec:hamiltonian-and-action}
Throughout this paper, we adopt units in which $\hbar = k_B = 1$.
The Hamiltonian density of a three-dimensional Fermi gas in the presence of Rashba-Dresselhaus
spin-orbit coupling and Rabi field is
\begin{equation}
{\cal H} ({\bf r})
=
{\cal H}_{k} ({\bf r}) + {\cal H}_{so} ({\bf r}) + {\cal H}_I ({\bf r})  - \mu n ({\bf r}).
\label{eqn:hamiltonian}
\end{equation}
The first term in Eq.~(\ref{eqn:hamiltonian}) is the kinetic energy,
\begin{eqnarray}
{\cal{H}}_k ({\bf r})
=
\sum_s
\psi^\dagger_s({\bf r})
\frac{{\hat {\bf k}}^2}{2m}
\psi_s ({\bf r})   ,
\end{eqnarray}
where ${\hat {\bf k}} = - i {\boldsymbol \nabla}$ is the momentum operator, 
$\psi_s({\bf r})$ is the fermion field at position ${\bf r}$ with (real or pseudo-)
spin $s$ and mass $m$. The second term is 
the spin-orbit interaction,
\begin{equation}
{\cal H}_{so} ({\bf r})
=
\sum_{s s^\prime}
\psi_{s}^\dagger ({\bf r})
\left[
{\bf H}_{so} ({\hat {\bf k}})
\right]_{s s^\prime}
\psi_{s^\prime} ({\bf r}), 
\end{equation}
with the spin-orbit coupling matrix in momentum $({\bf k})$ space being
\begin{equation}
{\bf H}_{so} (\hat {\bf k})
=
\frac{\kappa}{m} ( {\hat k}_x {\boldsymbol \sigma}_x + \eta {\hat k}_y {\boldsymbol \sigma}_y )
- \frac{\Omega_R}{2} \hspace{.5mm} {\boldsymbol \sigma}_z   ,
\end{equation}
where $({\boldsymbol \sigma}_x, {\boldsymbol \sigma}_y, {\boldsymbol \sigma}_z)$
are the Pauli matrices in spin space, $\kappa$ is the momentum transfer to the atoms in a
two-photon Raman process~\cite{spielman-2013a} or on a
radio frequency atom chip~\cite{spielman-2010}, $\eta$ is the anisotropy of the
Rashba-Dresselhaus field, and $\Omega_R$ is the Rabi frequency.
The third term is the two-body $s$-wave
contact interaction,
\begin{equation}
{\cal H}_I ({\bf r})
=
-  g \psi^\dagger_{\uparrow}
({\bf r}) \psi^\dagger_{\downarrow} ({\bf r}) \psi_{\downarrow} ({\bf r}) \psi_{\uparrow} ({\bf r})   ,
\label{eqn:interaction-hamiltonian}
\end{equation}
where $g > 0$ corresponds to a constant attraction between opposite spins.
Finally, $\mu$ is the chemical potential and
$n ({\bf r}) = \sum_s \psi^\dagger_s ({\bf r}) \psi_s  ({\bf r})$ is the local density.
While the general Rashba-Dresselhaus spin-orbit coupling is discussed
in Appendix~\ref{sec:appendix-A}, in what follows we focus on the more experimentally
relevant situation of equal Rashba and Dresselhaus couplings ($\eta = 0$).

Standard manipulations (see Appendix~\ref{sec:appendix-A}) lead to the Lagrangian density,
\begin{eqnarray}
{\cal L} ({\bf r, \tau}) & = & \frac{1}{2} \hspace{.5mm}
\Psi^\dagger ({\bf r},\tau) {\bf G}^{-1} (\hat{\bf k},\tau) \Psi ({\bf r},\tau)
+ \frac{1}{g} |\Delta ({\bf r},\tau)|^2 \label{eqn:lagrangian} \nonumber   \\
&& \hspace{5mm} + K (\hat{\bf k}) \delta ({\bf r} - {\bf r}^\prime)   ,   
\end{eqnarray}
where $\tau = it$ is the imaginary time,
$\Psi = (\psi_{\uparrow} \hspace{1mm} \psi_{\downarrow} \hspace{1mm} \psi^\dagger_{\uparrow} \hspace{1mm} \psi^\dagger_{\downarrow})^T$ is the Nambu spinor,
$K (\hat {\bf k}) =  {\hat {\bf k}}^2 /2 m - \mu$ is the kinetic energy operator with respect
to the chemical potential, and 
$
\Delta ({\bf r},\tau) = - g \langle \psi_\downarrow ({\bf r},\tau) \psi_\uparrow ({\bf r},\tau) \rangle
$ 
is the pairing field describing the formation of pairs of two fermions with opposite spins.
Note that $\mu$ includes the overall positive shift $\kappa^2/2m$ in the single-particle
kinetic energies due to spin-orbit coupling.  The inverse Green's operator
appearing in Eq.~(\ref{eqn:lagrangian}) is
\begin{eqnarray}
\label{eqn:greens-function}
{\bf G}^{-1} (\hat{\bf k}, \tau)
= 
\begin{pmatrix}			
\partial_\tau - K_\uparrow &- i\kappa {\hat k}_x/m & 0 & -\Delta \\
i\kappa {\hat k}_x/m & \partial_\tau - K_\downarrow & \Delta & 0 \\
0 & \Delta^* & \partial_\tau + K_\uparrow & -i\kappa {\hat k}_x/m \\
-\Delta^* & 0 & i\kappa {\hat k}_x/m & \partial_\tau + K_\downarrow
\end{pmatrix},   \nonumber\\
\end{eqnarray}
where
$ K_{{\uparrow,\downarrow}} = K (\hat {\bf k}) \mp \Omega_R/2, $ 
are the kinetic energy terms shifted by the Rabi coupling.

As noted above, a mean-field treatment of this Lagrangian fails to correctly
describe the superfluid critical temperature in the BEC regime.  However, the inclusion of
Gaussian fluctuations of $\Delta$ captures the effects of two-body bound states and
leads to a physical superfluid transition temperature.  It is to this task that we now turn.

\section{Thermodynamic Potential}
\label{sec:thermodynamic-potential}
The system's partition function may be expressed in terms of the functional integral,
\begin{equation}
{\cal{Z}} =
\int {\cal{D}} \Delta {\cal{D}}
 \Delta^*{\cal{D}} \Psi {\cal{D}} \Psi^\dagger \hspace{.5mm} e^{-{\cal{S}}}   ,
\end{equation}
where the Euclidean action is
\begin{equation}
\label{eqn:full-action}
{\cal S} = \int_0^\beta d\tau \int d^3 {\bf r} \hspace{.5mm} {\cal L}({\bf r}, \tau)   ,
\end{equation}
$\beta = 1/T$ is the inverse temperature, and the Lagrangian density
is given by Eq.~(\ref{eqn:lagrangian}).  Integrating over the fermion fields yields
the thermodynamic potential,
\begin{equation}
\Omega = -T \ln {\cal{Z}} = \Omega_0 + \Omega_F   ,
\end{equation}
where $\Omega_0 = -T \ln {\cal Z}_0 = T S_0$ is the mean-field (saddle-point) contribution,
for which $\Delta ({\bf r}, \tau) = \Delta_0$, 
and $\Omega_F = - T \ln {\cal Z}_F$ is the contribution arising from order-parameter fluctuations.
Detailed derivations of the thermodynamic potential for a general Rashba-Dresselhaus spin-orbit
coupling, as well as the associated order parameter and number equations,
are given in Appendices~\ref{sec:appendix-B}
and~\ref{sec:appendix-C}.
The contributions to the thermodynamic potential for the experimentally
relevant situation of equal Rashba-Dresselhaus spin-orbit coupling
are discussed below in Sec.~\ref{sec:saddle-point-approximation} at the mean-field and
in Sec.~\ref{sec:gaussian-fluctuations} at the Gaussian fluctuation level.

\subsection{Mean-Field Approximation}
\label{sec:saddle-point-approximation}

The mean-field, or saddle-point, term in the thermodynamic potential is
\begin{equation}
\Omega_0 
= 
V\frac{|\Delta_0|^2}{g} 
- 
\frac{T}{2} \sum_{{\bf k},j} 
\ln
\left[
1 + e^{-\beta E_{j} ({\bf k})} 
\right]
+ 
\sum_{\bf k} \xi_{\bf k},
\end{equation}
where $\xi_{\bf k} = \varepsilon_{\bf k}  - \mu$, $\varepsilon_{\bf k} = {\bf k}^2 /2 m$,
and the $E_{j} ({\bf k})$, with $j = \{1,2,3,4\}$, are the eigenvalues of the 
momentum space Nambu Hamiltonian matrix,
\begin{equation}
\label{eqn:nambu-hamiltonian}
{\bf H}_0 ({\bf k}) = {\boldsymbol \partial}_\tau -
{\bf G}^{-1} ({\bf k}, \tau) \vert_{\Delta = \Delta_0},
\end{equation}
where the operator ${\boldsymbol \partial}_{\tau} = {\bf I}   \partial_{\tau}$,
and ${\bf I}$ is the identity matrix. 
The first set of eigenvalues,
\begin{equation}
\label{eqn:quasiparticle-excitation-energy}
E_{1,2} ({\bf k})
=
\left[
\zeta^2_{\bf k}
\pm 2 \sqrt{E_{0,{\bf k}}^2 h_{\bf k}^2
- \left(\frac{\kappa k_x}{m} \right)^2 |\Delta_0|^2} \hspace{.5mm}
\right]^{1/2}
\end{equation}
describe quasiparticle excitations, with the plus $(+)$ associated with $E_1$ and the
minus $(-)$ with $E_2$.  The second set of eigenvalues,
$
E_{3,4} ({\bf k})
= 
- 
E_{2,1} ({\bf k})   ,
$
corresponds to quasiholes.
Further, $\zeta^2_{\bf k} = E^2_{0,{\bf k}} + h^2_{\bf k}$, where 
$
E_{0,{\bf k}} 
= \sqrt{
\xi^2_{\bf k} 
+ 
|\Delta_0|^2},
$
and $ h_{\bf k} = \sqrt{ (\kappa k_x/m)^2
 + \Omega^2_R / 4}$ is the magnitude of the combined spin-orbit and
Rabi couplings.

We express the two-body interaction parameter $g$ in terms of the renormalized
$s$-wave scattering length $a_s$ via the relation~\cite{sademelo-1993}
\begin{equation} 
\frac{1}{g} = - \frac{m}{4 \pi a_s} + \frac{1}{V} \sum_{\bf k} \frac{1}{2\varepsilon_{\bf k}}   .
\end{equation}
Note that $a_s$ is the $s$-wave scattering length {\it in the absence} of spin-orbit
and Rabi fields.  It is, of course, possible to express $g$, and all subsequent relations, 
in terms of a scattering length which is renormalized by the presence 
of the spin-orbit and Rabi fields~\cite{goldbart-2011, ozawa-2012}, but for both
simplicity and the sake of referring to the more experimentally accessible quantity,
we do not do so here.

The order-parameter equation is obtained from the saddle-point condition
$\delta \Omega_0 / \delta \Delta_0^* \vert_{T, V, \mu} = 0$, leading to
\begin{equation}
\label{eqn:saddle-point-order-parameter}
\frac{m}{4 \pi a_s} 
= 
\frac{1}{2 V} 
\sum_{\bf k} 
\bigg[ 
\frac{1}{\varepsilon_{\bf k}}
- 
A_{+} ({\bf k})
- 
\frac{\Omega^2_R}{4\xi_{\bf k} h_{\bf k}} 
\hspace{.5mm} 
A_{-} ({\bf k})
\bigg]   ,   
\end{equation}
where we introduced the notation
\begin{equation}
A_{\pm} ({\bf k})
=
\frac{1 - 2 n_{1}({\bf k})}{2 E_1 ({\bf k})}
\pm
\frac{1 - 2 n_{2}({\bf k})}{2 E_2 ({\bf k})}, 
\end{equation}
with 
$n_{j} ({\bf k})
= 
1 /
\left[
 e^{\beta E_{j }({\bf k})} + 1
\right]
$
being the Fermi function.  In addition, the particle number at the saddle point
$
N_0 
= 
- \partial \Omega_0 / \partial \mu \vert_{T, V},
$ 
is given by
\begin{equation}
\label{eqn:saddle-point-number}
N_0 
=  
\sum_{\bf k} 
\bigg \{ 
1 - \xi_{\bf k}
\bigg[ 
A_+ ({\bf k})
+ 
\frac{(\kappa k_x/m)^2}{\xi_{\bf k} h_{\bf k}} 
\hspace{.5mm} 
A_-  ({\bf k})
\bigg]  
\bigg \}.   
\end{equation}

The mean-field temperature $T_0$ is 
determined by solving Eq.~(\ref{eqn:saddle-point-order-parameter}) for the given $\mu$.
The corresponding number
of particles is given by Eq.~(\ref{eqn:saddle-point-number}). 
This mean-field treatment leads to a transition temperature 
$\sim e^{1/k_F a_s}$, where $k_F$ is the Fermi momentum.  This result gives the
correct transition temperature on the BCS limit; however, it is unphysical on the BEC regime
for $k_Fa_s \to 0$.  In order to find a physical result, we need to
include order-parameter fluctuations, which we now do.

\subsection{Gaussian Fluctuations}
\label{sec:gaussian-fluctuations}
In discussing Gaussian fluctuations, we concentrate on equal Rasha-Dresselhaus couplings,
leaving details for general Rashba-Dresselhaus coupling to Appendix~\ref{sec:appendix-C}.

To obtain the correct superfluid transition temperature in the BEC limit we must
include the physics of two-body bound states near the transition, as described by
the two-particle $T$-matrix~\cite{nozieres-1985, baym-2006}.
Accounting for all two-particle channels, 
the T-matrix calculation leads to a two-particle scattering amplitude 
$\Gamma$, where
\begin{equation}
\label{eqn:gamma}
\Gamma^{-1} ({\bf q},z) 
= 
\frac{m}{4 \pi a_s} 
- 
\frac{1}{2V} \sum_{\bf k} 
\bigg[ 
\frac{1}{~\varepsilon_{\bf k}} 
+ 
\sum^2_{i,j=1} 
\alpha_{ij} W_{ij} 
\bigg];  
\end{equation}
$z$ is the complex frequency and
\begin{equation}
\label{eq:Ws}
W_{ij}
=
\frac{1 - n_{i}({\bf k}) - n_{j}({\bf k+q})}{z - E_i({\bf k}) - E_j({\bf k +q}) }.
\end{equation}
At the superfluid phase boundary $\Delta_0 \to 0$, the eigenvalues
appearing in Eq.~(\ref{eq:Ws}) reduce to
$E_{1,2} ({\bf k}) = \left| |\xi_{\bf k}| \pm h_{\bf k} \right|$,
but it is straightforward to show that
ignoring the absolute values does not result in any change in either the
mean-field order parameter or number equation.
Meanwhile, the coefficients
\begin{eqnarray}
\alpha_{11} 
& = &  
\alpha_{22} 
= \vert
u_{\bf k} u_{{\bf k}+{\bf q}} 
- 
v_{\bf k} v^*_{{\bf k}+{\bf q}}
\vert^2   ,   \\
\alpha_{12} 
& = & 
\alpha_{21} 
= 
\vert 
u_{\bf k} v_{{\bf k}+{\bf q}} 
+ 
u_{{\bf k}+{\bf q}} v_{\bf k}
\vert^2   ,
\end{eqnarray}
are the coherence factors associated with  the quasi-particle amplitudes for $\Delta_0 = 0$:
\begin{equation}
u_{\bf k} 
= 
\sqrt{\frac{1}{2} 
\left(
1 + \frac{\Omega_R}{2h_{\bf k}} 
\right)},   
\hspace{5mm}   
v_{\bf k} 
= i\sqrt{\frac{1}{2} 
\left(1 - \frac{\Omega_R}{2h_{\bf k}} 
\right)}.
\end{equation}

The Gaussian fluctuation correction to the thermodynamic potential is
\begin{equation}
\Omega_F 
= 
- 
T 
\sum_{{\bf q},i q_n} 
\ln 
\left[ 
\beta
\Gamma ({\bf q},i q_n)/V
\right]   .
\end{equation}
over the entire BCS-to-BEC evolution. 
The fluctuation contribution to the particle number is therefore
$N_F = - \partial \Omega_F / \partial \mu \vert_{T, V}$, where
\begin{equation}
\label{eqn:fluctuation-number}
N_F
= 
\sum_{\bf q} 
\int^\infty_{-\infty} 
\frac{d \omega}{\pi} 
n_B (\omega) 
\left[
\frac{\partial \delta ({\bf q},\omega)}{\partial \mu}
-\frac{\partial \delta ({\bf q},0)}{\partial \mu}
\right]_{T,V},   
\end{equation}
with the phase shift $\delta ({\bf q},\omega)$
defined via the relation 
\begin{equation}
\Gamma ({\bf q},\omega \pm i \epsilon) 
= 
|\Gamma ({\bf q},\omega)| e^{\pm i \delta({\bf q},\omega)}.
\end{equation}
When two-body states are present, the fluctuation contribution can be written as
$N_F = N_{sc} + N_b$, where
\begin{equation}
\label{eqn:nscattering}
N_{sc} 
=  
\sum_{\bf q} 
\int^\infty_{\omega_{tp} ({\bf q})} 
\frac{d \omega}{\pi} 
n_B (\omega) 
\left[
\frac{\partial \delta ({\bf q},\omega)}{\partial \mu}
-\frac{\partial \delta ({\bf q},0)}{\partial \mu}
\right]_{T,V}  
\end{equation}
is the number of particles in scattering states, 
and $\omega_{tp} ({\bf q})$ is the two-particle continuum threshold 
corresponding to the branch point
of $\Gamma^{-1} ({\bf q},z)$~\cite{nozieres-1985, pethick-2011},
\begin{equation}
\label{eqn:nbound}
N_b  
=
2 
\sum_{\bf q} n_B (E_{bs} ({\bf q}) - 2\mu),   
\end{equation}
is the number of fermions in bound states, where
$n_B (\omega) = 1 / (e^{\beta \omega} - 1)$ is the Bose distribution function, and
$E_{bs} ({\bf q})$ is
the energy of the bound states obtained from $\Gamma^{-1}({\bf q}, z = E - 2\mu) = 0$,
corresponding to a pole in the scattering amplitude $\Gamma ({\bf q}, z)$.
In the limit of large and negative fermion chemical potential,
the system becomes non-degenerate and $\Gamma^{-1}({\bf q}, z) = 0$ 
becomes the exact eigenvalue equation for the two-body bound state
in the presence of spin-orbit and Rabi coupling~\cite{doga-2016}.
The total number of fermions, as a function of $\mu$, thus becomes 
\begin{equation}
\label{eqn:number-total}
N = N_0 + N_F,
\end{equation}
where $N_0$ is
given in Eq.~(\ref{eqn:saddle-point-number}) and $N_F$ is the 
sum of $N_{sc}$ and $N_b$, as
discussed above~\cite{nozieres-1985, sademelo-1993}.

\section{Critical Temperature}
\label{sec:critical-temperature}

We calculate numerically the transition temperature $T_c$ between
the normal and uniform superfluid states, as a function of
the scattering parameter $1/k_Fa_s$, by simultaneously solving the order parameter
and number equations (\ref{eqn:saddle-point-order-parameter})  
and~(\ref{eqn:number-total}).  The solutions correspond to the minima of the free energy, 
${\cal F} = \Omega + \mu N$.  We do not discuss the cases of Fulde-Ferrell~\cite{fulde-1964}
or Larkin-Ovchinnikov~\cite{larkin-1965} nonuniform superfluid phases since they only exist over
a very narrow region of the phase diagram deep
in the BCS regime~\cite{fulde-1964, larkin-1965}, which is not
experimentally accessible for ultracold fermions.

\begin{figure}[tb]
\centering{}
\includegraphics[width=0.48\textwidth,height=0.35\textwidth]{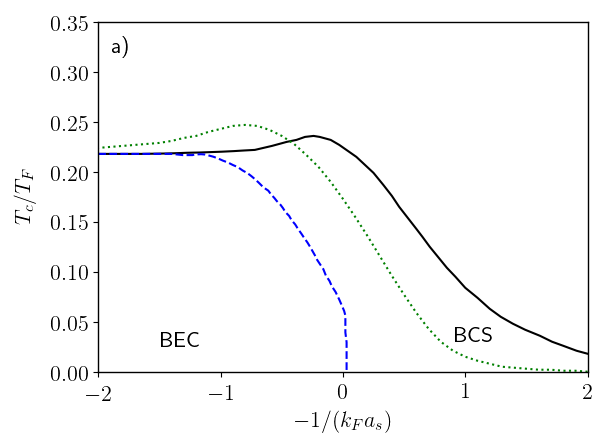}
\includegraphics[width=0.48\textwidth,height=0.35\textwidth]{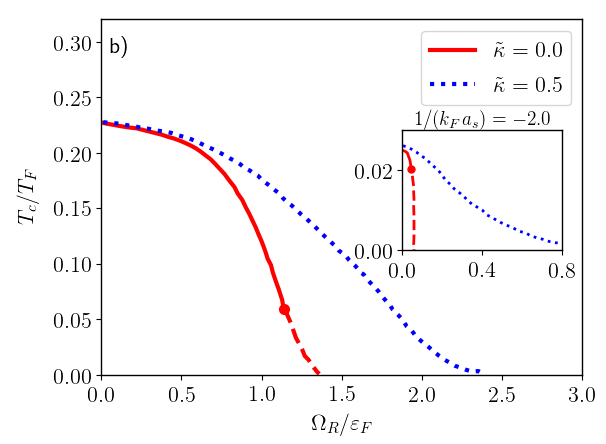}
\caption{
(Color online)
(a) The superfluid transition temperature $T_c/T_F$, where $T_F$ is the
Fermi temperature, vs
the scattering parameter $1/k_Fa_s$
for equal Rashba-Dresselhaus spin-orbit coupling
and two different Rabi coupling strengths, $\Omega_R = 0$ and $\varepsilon_F$.   
For $\Omega_R = 0$ [solid (black) curve], $T_c$ is the same as for
zero spin-orbit coupling
since the equal spin-orbit field can be gauged away.
The dashed (blue) line shows $T_c$ for zero spin-orbit coupling,
with $\Omega_R = \varepsilon_F$, while the dotted (green) line shows
$T_c$ for $\Omega_R = \varepsilon_F$ and $\tilde{\kappa} = \kappa/k_F = 0.5$.
(b) $T_c$ is drawn at unitarity, 
$1/k_F a_s = 0$, and in the inset at $1/k_F a_s = -2.0$, 
as a function of $\widetilde\Omega_R = \Omega_R/\varepsilon_F$.  
The solid (red) curves represent
$\tilde\kappa = 0$ and the dotted (blue) curves represent $\tilde\kappa = 0.5$.
Across the dotted (red) curves, the phase transition is first order.   
}
\label{fig:one}
\end{figure}

Figure~\ref{fig:one}, in which we scale temperatures by the Fermi temperature
$T_F = k_F^2/2m$, shows the effects of spin-orbit 
and Rabi couplings on $T_c$.  
The solid (black) line in Fig.~\ref{fig:one}(a)
shows $T_c$ versus $1/k_F a_s$ for zero Rabi coupling $(\Omega_R = 0)$
and zero spin-orbit coupling $\kappa$.
If $\Omega_R = 0$, the spin-orbit coupling can be removed by a
simple gauge transformation, and thus plays no role.
In this situation, the pairing is purely $s$-wave.
The dashed (blue) line shows $T_c$ for $\Omega_R \ne 0$, with vanishing
equal Rashba-Dresselhaus spin-orbit coupling. We see that for fixed interaction strength,
the pair-breaking effect of the Rabi coupling  suppresses superfluidity,
compared with $\Omega_R = 0$;  the Rabi field here plays the pair-breaking role
of the Zeeman field in an superconductor.

With both spin-orbit and Rabi couplings present, the two-particle pairing is no longer
purely singlet $s$-wave, but obtains a triplet $p$-wave component;
the admixture stabilizes the superfluid phase, as
shown by the dotted (green) line.
The latter curve shows that  in the BEC regime with large positive
$1/k_F a_s$, the superfluid transition temperature is enhanced by the presence of spin-orbit
and Rabi couplings, a consequence of the reduction 
of the bosonic effective mass in the $x$ direction below $2m$.
However, for sufficiently large $\Omega_R$, the geometric mean bosonic mass 
$M_B$ increases above $2 m$ and 
$T_c$ decreases.
This renormalization of the mass of the bosons can
be traced back to a change in the energy dispersion of the fermions
when both spin-orbit coupling and Rabi fields are present.

Figure~\ref{fig:one}(b) shows $T_c$ versus $\Omega_R$ 
for fixed $1/k_Fa_s$, both with and without
equal Rashba-Dresselhaus spin-orbit coupling at $\kappa = 0.5 k_F$.
When both $\kappa$ and $T$ are zero, superfluidity is destroyed 
at a critical value of $\Omega_R$ corresponding to the 
Clogston limit~\cite{clogston-1962}.  
At low temperature, the phase transition to the normal state is first order 
because the Rabi coupling is sufficiently large to 
break singlet Cooper pairs. 
However, at higher temperatures the singlet $s$-wave superfluid starts 
to become polarized by thermally excited quasiparticles that produce 
a paramagnetic response.   Thus, above the characteristic temperature 
indicated by the large (red) dots, the transition becomes second order,
as pointed out by Sarma~\cite{sarma-1963}.
The change in the transition order occurs not only for $\kappa = 0$,
but also for nonzero values of $\kappa$ both in the BCS regime and near
unitarity, depending on the choice of parameters, as illustrated in Fig.~\ref{fig:one}(b).

The critical temperature
for $\kappa \ne0$ vanishes only asymptotically  in the limit of large $\Omega_R$.   
We note that for $\Omega_R = E_F$ and $\kappa = 0$, the transition from
the superfluid to the normal state is continuous at unitarity, but very close
to a discontinuous transition.  In the range $1.05 \lesssim \Omega_R/E_F \lesssim 1.10$,
numerical uncertainties as $\kappa\to 0$ prevent 
us from predicting exactly whether the transition 
at unitarity is continuous or discontinuous.

\begin{figure}[tb]
\centering{}
\includegraphics[width=0.48\textwidth,height=0.35\textwidth]{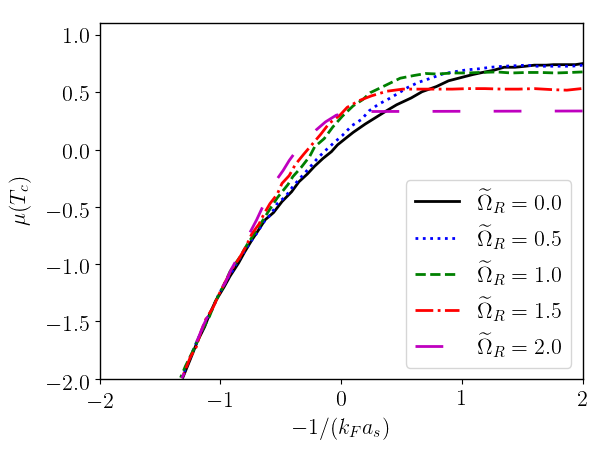}
\caption{
(Color online) Chemical potential at the superfluid critical temperature
($T_c$) for $\tilde{\kappa} = \kappa / k_F= 0.5$ and various Rabi fields,
${\widetilde\Omega}_R = \Omega_R / \varepsilon_F$.}
\label{fig:two}
\end{figure}

Figure~\ref{fig:two} shows $\mu (T_c)$ for fixed spin-orbit coupling and
several Rabi couplings.
The solid (black) curve, which represents the situation in which no Rabi field is present,
is equivalent to the situation in which spin-orbit coupling is also absent, as noted in the
discussion of Fig.~\ref{fig:one}.  It is evident that while the Rabi field reduces
the chemical potential in the BCS limit, it also shifts the onset of the
system's evolution to the BEC limit to larger inverse scattering lengths,
and produces a non-monotonic behavior of $\mu (T_c)$ near unitarity.

\begin{figure}[tb]
\includegraphics[width = 0.48\textwidth,height=0.35\textwidth] {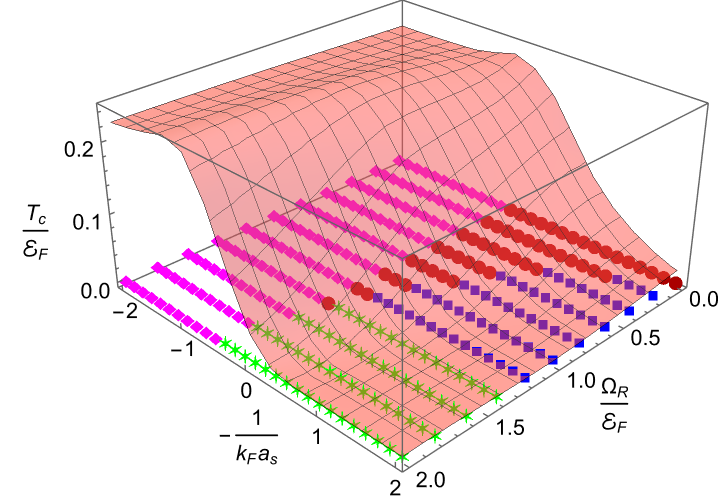}
\caption{
  (Color online)
Phase diagram of critical temperature $T_c/T_F$ vs $1/k_F a_s$ and $\Omega_R/\varepsilon_F$
for equal Rashba-Dresselhaus coupling $\kappa/k_F = 0.5$.
The finite-temperature uniform superfluid phases reflect those at $T = 0$
shown in the background. These phases are distinguished by the number of rings (line nodes) 
in the quasiparticle excitation spectrum [i.e., where $E_2 ({\bf k})= 0$] and type of gap:
(1) direct gapped superfluid with zero rings (magenta diamonds), 
(2) indirect gapped superfluid with zero rings (red circles), 
(3) gapless superfluid with two rings (blue square), and 
(4) gapless one-ring superfluid (green stars).}
\label{fig:three} 
\end{figure}

Figure~\ref{fig:three} shows $T_c$ for equal Rashba-Dresselhaus
coupling $\kappa = 0.5 k_F$, as a function of Rabi field and scattering 
parameter. We also superpose the zero-temperature phase diagram to illustrate
the different superfluid ground states of this system. According to the 
zeros of the lowest quasiparticle energy $E_2 ({\bf k})$, the uniform 
superfluid phases that emerge are~\cite{seo-2012a} 
direct gapped with zero rings (line nodes), 
indirectly gapped with zero rings, 
gapless with one ring, 
and gapless with two rings.

\begin{figure}[tb]
\includegraphics[width = 0.48\textwidth,height=0.35\textwidth]{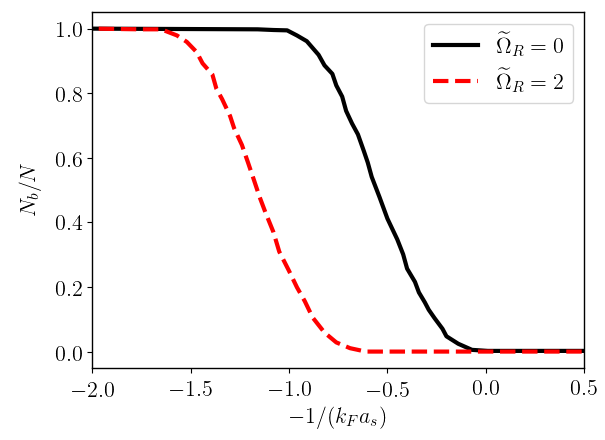}
\caption{
(Color online) 
Fractional number $N_b/N$ of bound fermions
as a function of the interaction parameter $1 / k_F a_s$, 
for equal Rashba-Dresselhaus coupling $\kappa / k_F = 0.5$
and Rabi frequencies
${\widetilde \Omega}_R =  \Omega_R/ \varepsilon_F = 0$
(black solid line)
and
${\widetilde \Omega}_R = \Omega_R/ \varepsilon_F = 2$
(red dot-dashed line).
}
\label{fig:four}
\end{figure}

Figure~\ref{fig:four} shows the fractional number $N_b/N$ of bound 
fermions at $T_c$ as a function of $1/k_F a_s$
for two sets of external fields.  
In the BCS (BEC) regime, the relative contribution to $N$ is dominated 
by unbound (bound) fermions. The main effect of 
spin-orbit and Rabi fields on $N_b/N$ is to shift the location 
where the two-body bound states emerge. For fixed spin-orbit coupling (Rabi field)
and increasing Rabi field (spin-orbit coupling), two-body bound states emerge at 
larger (smaller) scattering parameters. These shifts are in
agreement with the calculated shifts in binding energies of Feshbach
molecules in the presence of equal Rashba-Dresselhaus spin-orbit coupling
and Rabi fields~\cite{doga-2016}.

\section{Ginzburg-Landau Theory}
\label{sec:ginzburg-landau-theory}
To further elucidate the effects of fluctuations on the 
order of the superfluid transition, as well as to  
assess the impact of spin-orbit and Rabi couplings near the 
critical temperature, we now derive the 
Ginzburg-Landau description of the free energy near the transition.
In the limit of small order parameter,
the fluctuation action ${\cal S}_F$ can be expanded in powers of the  
order parameter $\Delta ({q})$ beyond Gaussian order.
The expansion of ${\cal S}_F$ to quartic order is sufficient to describe the
continuous (second-order) transition in $T_c$ versus $1/k_F a_s$ in the absence
of a Rabi field~\cite{sademelo-1993}.  However, to correctly describe
the first-order transition~\cite{clogston-1962, sarma-1963} at low
temperature (Fig.~\ref{fig:one}), it is necessary to
expand the free energy to sixth order in $\Delta$.

The quadratic (Gaussian-order) term in the action is
\begin{eqnarray}
 {\cal S}_G
& = & \beta V \sum_q \frac{|\Delta_q|^2}{\Gamma ({\bf q}, z)}.
\end{eqnarray}
For an order parameter varying slowly in space and time, we may expand $\Gamma^{-1}$ as
\begin{equation}
\Gamma^{-1} ({\bf q},z) 
= a  + \sum_\ell c_\ell \frac{q^2_\ell}{2 m}  - d_0 z  + \cdots, 
\label{eq:Gamma_expansion}
\end{equation}
with the sum over $ \ell = \{x, y ,z \}$.
The full result, as a functional of $\Delta ( {\bf r},\tau)$, has the form
\begin{eqnarray}
 {\cal S}_F
 & = & \int_0^\beta d\tau \int d^3 {\bf r}
 \Big( d_0 \Delta^* \frac{\partial}{\partial \tau}\Delta+a|\Delta|^2 \nonumber\\
&& + \sum_\ell c_\ell \frac{ | \nabla_\ell \Delta|^2}{2 m} + 
\frac{b}2 |\Delta|^4 + 
\frac{f}3 |\Delta|^6 \Big).
\end{eqnarray}
The full time-dependent Ginzburg-Landau action describes 
systems in and near equilibrium (e.g., with collective modes).
The imaginary part of $d_0$ measures the non-conservation
of $|\Delta|^2$ in time (i.e., the Cooper pair lifetime).
Details of the derivation of ${\cal S}_F$ are found in Appendix~\ref{sec:appendix-D}.

We are interested in systems at 
thermodynamic equilibrium, where the order parameter is 
independent of time, that is, $\Delta ({\bf r}, \tau) = \Delta ({\bf r})$.
In this situation, minimizing the
free energy $T{\cal S}_F$ with respect to $\Delta^*$
yields the Ginzburg-Landau equation,
\begin{equation}
\label{eqn:tdgl}
\left(
- \sum_\ell  c_\ell \frac{\nabla^2_\ell}{2 m}
+ b \vert \Delta ({\bf r}) \vert^2 + f \vert \Delta ({\bf r}) \vert^4 + a
\right) \Delta ({\bf r}) = 0 .   
\end{equation}
For $b>0$, the system undergoes a continuous 
phase transition when $a$ changes sign.   
However, when $b<0$, the system is unstable in the absence of $f$.  For $b<0$ and 
 $a > 0$, a first-order phase transition
occurs when $3 b^2 = 16 a f$.
Positive $f$ stabilizes the system even when $b < 0$.

In the BEC regime, where $d_0$ is purely real, we define an effective bosonic 
wave function $\Psi ({\bf r}) = \sqrt{d_0} \Delta ({\bf r})$ to recast Eq.~(\ref{eqn:tdgl}) 
in the form of the Gross-Pitaevskii equation for a dilute Bose gas,
\begin{equation}
\label{eqn:gp}
\left(
- \sum_\ell \frac{\nabla^2_\ell}{2 M_{\ell}} 
+ U_{2} \vert \Psi ({\bf r}) \vert^2 
+U_{3} \vert \Psi ({\bf r}) \vert^4 
-\mu_B
\right) \Psi ({\bf r}) = 0.   
\end{equation}
Here, $\mu_B = - a / d_0$ is the bosonic chemical potential,
$M_{\ell} = m (d_0 / c_\ell)$ are the anisotropic bosonic masses, 
and $U_{2} = b / d_0^2$ and $U_{3} = f/d_0^3$ represent contact
interactions of two and three bosons.  In the BEC regime, these terms are   
always positive, leading to a dilute gas of stable bosons.   
The boson chemical potential $\mu_B$ is $\approx 2 \mu + E_b <  0$,
where $E_b = - E_{bs} ({\bf q} = {\bf 0})$ is the two-body binding  energy in the presence
of spin-orbit coupling and Rabi frequency, obtained from the condition 
$\Gamma^{-1} ({\bf q}, E - 2\mu)= 0$, discussed earlier. 

The anisotropy of the effective bosonic masses, $M_{x} \ne M_{y} = M_{z} \equiv M_\perp$,
stems from the anisotropy of the equal Rashba-Dresselhaus spin-orbit coupling, which together 
with the Rabi coupling modifies the dispersion of the constituent 
fermions along the $x$ direction.  In the limit $k_Fa_s \ll 1$, the 
many-body effective masses reduce to those obtained by expanding the 
two-body binding energy, 
$E_{bs}({\bf q}) \approx-E_b + \sum_\ell q^2_\ell / 2 M_{\ell},$
and agree with known results~\cite{doga-2016}.
However, for $1/k_F a_s \lesssim 2$, many-body and thermal effects produce 
deviations from the two-body result. 

In the absence of two- and three-body boson-boson interactions, $U_2$ and $U_3$, we
directly obtain an analytic expression for $T_c$ in the 
Bose limit from Eq.~(\ref{eqn:nbound}),
\begin{equation}
  T_c = \frac{2\pi}{M_B}\left(\frac{n_B}{\zeta(3/2)}\right)^{2/3}, 
\end{equation}
with $M_B= (M_xM_\perp^2)^{1/3}$,
by noting that $\mu_B = 0$ or 
$E_{bs} ({\bf q} = {\bf 0}) - 2\mu = 0$, 
and using the condition that  $n_B \simeq n/2$  [with corrections exponentially
small in $(1/k_Fa_s)^2$], where $n_B$ is the
density of bosons. 
In the BEC regime, the results shown in Fig.~\ref{fig:one}
include the effects of the mass anisotropy, but do not include the effects 
of boson-boson interactions. 

To account for boson-boson interactions, we adopt the Hamiltonian of
 Eq.~(\ref{eqn:gp}) with $U_2 \ne 0$, but with $U_3 = 0$, and 
 apply the method developed in Ref.~\cite{baym-1999} to show
that these interactions further increase $T_{BEC}$ to 
\begin{equation}
T_c (a_B) = (1 + \gamma) T_{BEC},
\end{equation}
where $\gamma = \lambda n_B^{1/3} a_B$. 
Here, $a_{B}$ is the $s$-wave boson-boson scattering 
length,  $\lambda$ is a dimensionless constant $\sim 1$,
and we use the relation $U_2 = 4\pi a_B/M_B$. 
Since 
$n_B =  k_F^3/6 \pi^2$ 
and the boson-boson scattering length is
$a_B = U_2 M_B/4\pi$, we have
$\gamma = {\tilde \lambda} {\widetilde M}_B {\widetilde U}_2,$
where 
$
{\widetilde M}_B 
= 
M_B/2m,
$ 
$
{\widetilde U}_2 =
U_2 k_F^3/\varepsilon_F   ,
$
and 
$
{\tilde \lambda}
= 
\lambda
/
4(6\pi^5)^{1/3}
\approx 
\lambda/50 .
$
For fixed $1/k_F a_s$, $T_c$ is enhanced by the spin-orbit field, a $\Omega_R$-dependent 
decrease in the effective boson mass $M_B$ ($\sim$10-15\%), 
as well as a stabilizing boson-boson repulsion $U_2$ ($\sim$2-3\%),
for the parameters used in Fig.~\ref{fig:one}.

In closing our discussion of the strongly bound BEC limit, we note that
in the absence of spin-orbit coupling, a Gaussian-order calculation of the two-boson scattering
length yields the erroneous Born approximation result $a_B = 2 a_s$.  However, an analysis
of the $T$-matrix beyond Gaussian order, which includes the effects of two-body bound states,
obtains the correct result $a_B = 0.6 a_s$ at very low densities~\cite{iskin-2008} and agrees
with four-body calculations~\cite{petrov-2005}.
The same method can be used to estimate $U_2$ or $a_B$ beyond the Born approximation discussed
above.  Nevertheless, while the precise quantitative relation between $a_B$ and $a_s$ in the
presence of spin-orbit coupling is yet unknown, the trend of increasing $T_c$ due to spin-orbit coupling
has been clearly shown.

\section{Comparison to earlier work}
\label{sec:comparison-to-earlier-work}
In this section, we briefly compare our results with earlier investigations of different types
of theoretically motivated
spin-orbit couplings, worked in different dimensions, or at zero temperature.
Our results focus mainly on an analysis of the critical superfluid temperature 
and the effects thereon of order-parameter fluctuations for a three-dimensional Fermi gas
in the presence of equal Rashba-Dresselhaus spin-orbit coupling and Rabi fields.
The appendices consider the more general situation of arbitrary Rashba and Dresselhaus
components.

Several works have analyzed the effects of spin-orbit-coupled fermions
in three dimensions at zero
temperature~\cite{zhang-2011, zhai-2011, pu-2011, han-2012a, seo-2012a, seo-2012b, shenoy-2011, feng-2020, dellanna-2011, dellanna-2012}.
While some authors have described the situation of
Rashba-only couplings~\cite{shenoy-2011, zhang-2011, zhai-2011,pu-2011}, others
have assessed the case of equal Rashba and Dresselhaus
components~\cite{seo-2012a, seo-2012b} or a general
mixture of the two~\cite{han-2012a}.  It has been demonstrated that
in the absence of a Rabi field, the zero-temperature evolution from
BCS to BEC superfluidity
is a crossover for $s$-wave systems, not only for Rashba-only
couplings~\cite{shenoy-2011, zhang-2011, zhai-2011, pu-2011, han-2012a}, but also for
arbitrary Rashba and Dresselhaus components~\cite{han-2012a}.
This result directly follows from the fact that the quasiparticle excitation
spectrum remains fully gapped throughout the evolution.

In contrast, the addition of a Rabi field gives rise to topological phase transitions
for Rashba-only
couplings~\cite{zhang-2011} and
equal Rashba and Dresselhaus components~\cite{seo-2012a, seo-2012b},
a situation which certainly persists for general Rashba-Dresselhaus couplings.
The simultaneous presence
of a general Rashba-Dresselhaus spin-orbit coupling and Rabi fields leads to
a qualitative change in the
quasiparticle excitation spectrum and to the emergence of topological superfluid
phases~\cite{zhang-2011, seo-2012a, seo-2012b}.
Two-dimensional systems have also been investigated at zero temperature,
where topological phase
transitions have been identified for Rashba-only~\cite{tewari-2011} and
equal Rashba-Dresselhaus~\cite{han-2012b} couplings, in the presence of a Rabi field.

While early papers in this field focused mainly on the zero-temperature limit,
progress toward finite-temperature theories was made first in
two dimensions~\cite{devreese-2014, devreese-2015} and later in
three dimensions~\cite{shenoy-2015, ohashi-2016, levin-2015}.
The effects of a general Rashba-Dresselhaus spin-orbit coupling and Rabi field on
the Berezenskii-Kosterlitz-Thouless transition were thoroughly investigated for
two-dimensional Fermi gases
at finite temperatures~\cite{devreese-2014, devreese-2015},
including both Rashba-only and equal Rashba-Dresselhaus spin-orbit
couplings as examples.

The superfluid critical temperature in three dimensions was investigated using
a spherical (3D) spin-orbit coupling $\lambda {\bf k} \cdot {\boldsymbol \sigma}$
in the absence of a Rabi field~\cite{shenoy-2015, ohashi-2016}, and also for
Rashba-only (2D) couplings in the presence of a Rabi field~\cite{levin-2015}.
In a recent review article~\cite{dellanna-2021}, the critical temperature throughout the
BCS-BEC evolution was discussed both in the absence~\cite{sademelo-1993}
and presence~\cite{powell-2017} of Rashba-Dresselhaus spin-orbit coupling.
In Secs. 5 and 6 of this review, the authors describe the same method and
expressions we obtained in our earlier preliminary work~\cite{powell-2017} for the analytical relations required
to obtain the critical temperature at the Gaussian order; they include, however, only the contribution 
of bound states discussed earlier in the literature for 
Rashba-only spin-orbit coupling without Rabi fields~\cite{zhai-2011}.
In contrast, here we develop a complete Gaussian theory to
compute the superfluid critical temperature
of a three-dimensional Fermi gas in the presence
of both a general Rashba-Dresselhaus (2D) spin-orbit coupling and Rabi fields.
We focus our numerical calculations on the specific situation of
equal Rashba-Dresselhaus components, which
is easier to achieve experimentally in the context of ultracold atoms. Our key results, already
announced in our earlier work~\cite{powell-2017}, include the contributions of bound and
scattering states at the Gaussian level. As seen in Fig.~\ref{fig:four} of this present paper, there is a wide
region of interaction parameters for which the contribution of scattering states cannot be
neglected. Furthermore, unlike previous work~\cite{shenoy-2015, ohashi-2016, levin-2015,dellanna-2021},
we provide a comprehensive analysis of the Ginzburg-Landau fluctuation
theory and include the effects of boson-boson interactions on the superfluid critical temperature
in the BEC regime.

\section{Conclusion}
\label{sec:conclusions}
We have evaluated the superfluid critical temperature throughout the BCS-to-BEC evolution of
three-dimensional Fermi gases in the presence of equal Rashba-Dresselhaus spin-orbit couplings, 
Rabi fields, and tunable $s$-wave interactions.  Furthermore, 
we have developed the Ginzburg-Landau theory up to sixth power 
in the order parameter to elucidate the origin of  
first-order phase transitions when the spin-orbit field is absent and the Rabi field
is sufficiently large.  Lastly, in the appendices,
we have presented the finite-temperature theory of $s$-wave interacting fermions
in the presence of a generic Rashba-Dresselhaus coupling and external
Rabi fields, as well as the corresponding time-dependent Ginzburd-Landau theory
near the superfluid critical temperature.

\acknowledgments{
We  thank I. B. Spielman for discussions.
The research of P.D.P. was supported in part by  
NSF Grant No. PHY1305891 and that of G.B. by NSF Grants No. PHY1305891 and No. PHY1714042.  
Both G.B. and C.A.R. SdM. thank the Aspen Center for Physics, 
supported by NSF Grants No. PHY1066292 and No. PHY1607611, where 
part of this work was done.  This work was performed under the auspices of the 
U.S. Department of Energy by Lawrence Livermore National Laboratory under 
Contract No. DE-AC52-07NA27344.
}

\appendix
\section{Hamiltonian and effective Lagrangian for general Rashba-Dresselhaus
spin-orbit coupling}
\label{sec:appendix-A}

In this appendix, we  consider a larger class of spin-coupled fermions in three dimensions
with a general Rashba-Dresselhaus (GRD) coupling. The Hamiltoninan density for
equal Rashba-Dresselhaus (ERD) discussed in Sec.~\ref{sec:hamiltonian-and-action}
is a particular case of the general Rashba-Dresselhaus Hamiltonian density,
\begin{equation}
{\cal H} ({\bf r}) = {\cal H}_0 ({\bf r}) + {\cal H}_{so} ({\bf r}) + {\cal H}_I ({\bf r}).
\end{equation}
Adopting units in which $\hbar = k_B = 1$, the independent-particle Hamiltonian density
without spin-orbit coupling is
\begin{equation}
{\cal H}_0 ({\bf r}) 
= 
\sum_\alpha 
\left( 
\frac{|\nabla \psi_\alpha ({\bf r})|^2}{2 m_\alpha} 
- \mu_\alpha \psi^\dagger_\alpha ({\bf r}) \psi_\alpha ({\bf r}) 
\right),
\end{equation}
where $\psi_\alpha$, $m_\alpha$, and $\mu_\alpha$ are the fermion field operator, 
mass, and chemical potentials for internal state $\alpha$, respectively.  
The spin-orbit Hamiltonian can be written as 
\begin{equation}
{\cal H}_{so} ({\bf r}) 
= 
- 
\sum_{i \alpha \beta} 
\psi^\dagger_\alpha ({\bf r}) 
\sigma_{i,\alpha \beta} 
h_i ({\bf r}) 
\psi_\beta ({\bf r}),
\end{equation}
where the ${\boldsymbol \sigma}_i$ are the Pauli matrices in isospin (internal state) 
space and ${\bf h} = (h_x, h_y, h_z)$ includes both the spin-orbit coupling 
and Zeeman fields.  Finally, we consider a two-body $s$-wave contact 
interaction,
\begin{equation}
{\cal H}_I ({\bf r}) 
= 
- 
g 
\psi^\dagger_{\uparrow} ({\bf r}) 
\psi^\dagger_{\downarrow} ({\bf r}) 
\psi_{\downarrow} ({\bf r}) 
\psi_{\uparrow} ({\bf r}),
\end{equation}
where $g > 0$ corresponds to an attractive interaction.

By introducing the pairing field 
$
\Delta ({\bf r},\tau) 
= 
- 
g 
\langle 
\psi_\downarrow ({\bf r},\tau) 
\psi_\uparrow ({\bf r},\tau)
\rangle,
$ 
we remove the quartic interaction and obtain the Lagrangian density,
\begin{eqnarray}
\label{eqn:lagrangian-density-appendix}
{\cal L} ({\bf r, \tau}) & = & 
\frac{1}{2} 
\hspace{.5mm} 
\Psi^\dagger ({\bf r},\tau) 
{\bf G}^{-1} (\hat{\bf k},\tau) 
\Psi ({\bf r},\tau) 
+ \frac{|\Delta ({\bf r},\tau)|^2}{g}   
\nonumber   
\\
&& \hspace{5mm} 
+ 
\widetilde{K}_+ (\hat {\bf k}) \delta ({\bf r} - {\bf r}^\prime),   
\end{eqnarray}
where we introduced the momentum operator $\hat {\bf k} = -i {\bf \nabla}$,
the Nambu spinor 
$
\Psi 
= 
(\psi_{\uparrow} 
\hspace{1mm} 
\psi_{\downarrow} 
\hspace{1mm} 
\psi^\dagger_{\uparrow} 
\hspace{1mm} 
\psi^\dagger_{\downarrow})^T
$,
and defined 
$
\widetilde{K}_\pm 
= 
(\widetilde{K}_{\uparrow} 
\pm 
\widetilde{K}_{\downarrow})
/ 
2.
$
Here,  
$
\widetilde{K}_{\uparrow} 
= 
K_{\uparrow} 
- 
h_z,
$ 
and 
$
\widetilde{K}_{\downarrow} 
= 
K_{\downarrow} 
+ 
h_z,
$  
with 
$
K_\alpha  (\hat{\bf k})
= 
{\hat {\bf k}}^2 / (2 m_\alpha) 
- 
\mu_\alpha
$ 
being the kinetic energy operator of internal state $\alpha$ with respect to its chemical potential. 
Lastly, the inverse Green's operator appearing in Eq.~(\ref{eqn:lagrangian-density-appendix}) is
\begin{eqnarray}
 \label{eqn:greens-function-appendix}
{\bf G}^{-1} (\hat{\bf k},\tau) 
= 
\begin{pmatrix}			
\partial_\tau - \widetilde{K}_\uparrow & h^*_\perp & 0 & -\Delta \\
h_\perp & \partial_\tau - \widetilde{K}_\downarrow & \Delta & 0 \\
0 & \Delta^* & \partial_\tau + \widetilde{K}_\uparrow & -h_\perp \\
-\Delta^* & 0 & -h^*_\perp & \partial_\tau + \widetilde{K}_\downarrow
\end{pmatrix}, \nonumber\\  
\end{eqnarray}
where $h_\perp (\hat{\bf k}) =  h_x (\hat{\bf k})+ i h_y (\hat{\bf k})$ plays the role of the
spin-orbit coupling, and $h_z$ is the Zeeman field along the $z$ direction.

To make progress, we expand the order parameter
about its saddle-point (mean-field) value $\Delta_0$ by writing 
$
\Delta ({\bf r},\tau) 
= \Delta_0 
+ 
\eta ({\bf r},\tau).
$  
Next, we integrate over the fermionic fields and use the decomposition
$
{\bf G}^{-1} (\hat {\bf k},\tau) 
= 
{\bf G}_0^{-1} (\hat{\bf k},\tau) 
+ 
{\bf G}_F^{-1} (\hat{\bf k},\tau),
$
where  ${\bf G}_0^{-1} (\hat{\bf k}, \tau)$ is the mean-field Green's operator, given
by Eq.~(\ref{eqn:greens-function-appendix}) 
with $\Delta ({\bf r},\tau) = \Delta_0$, and ${\bf G}_F^{-1} (\hat{\bf k},\tau)$ is the
contribution to the inverse Green's operator arising from fluctuations.
These steps yield the saddle-point Lagrangian density,
\begin{equation}
\label{eqn:saddle-point-lagrangian-appendix}
{\cal L}_0 ({\bf r}, \tau) = 
- 
\frac{T}{2 V} 
\hspace{.5mm} 
{\rm Tr} \ln 
(\beta {\bf G}^{-1}_0) 
+ 
\frac{|\Delta_0|^2}{g} 
+ 
\widetilde{K}_+ (\hat{\bf k}) 
\delta({\bf r} - {\bf r}^\prime),  
\end{equation}
and the fluctuation contribution,
\begin{equation}
{\cal L}_F ({\bf r}, \tau) =  
- \frac{T}{2 V} 
\hspace{.5mm} 
{\rm Tr} \ln ({\bf I} + {\bf G}_0 {\bf G}^{-1}_F) 
+
\Lambda ({\bf r}, \tau) 
+
\frac{|\eta ({\bf r},\tau)|^2}{g},   
\label{eq:L_fluc}
\end{equation}
resulting in the effective Lagrangian density 
$
{\cal L}_{\rm eff} ({\bf r}, \tau)
=
{\cal L}_{0} ({\bf r}, \tau)
+
{\cal L}_{F} ({\bf r}, \tau).
$ 
In the expressions above, we work in a volume $V$ and take traces over 
both discrete and continuous indices. Notice that the term 
$
\Lambda ({\bf r}, \tau) 
= 
\left[
\Delta_0 \eta^* ({\bf r},\tau) 
+ 
\Delta_0^* \eta ({\bf r},\tau) 
\right]/g
$
in the fluctuation Lagrangian cancels out the linear terms in 
$\eta$ and $\eta^*$ when the logarithm is expanded, due to the saddle 
point condition
\begin{equation}
\frac{\delta S_{0}}{\delta \Delta_0^*} = 0,
\end{equation}
where $S_0 = \int_0^\beta {d\tau} d^3{\bf r} {\cal L}_0 ({\bf r}, \tau)$ is the saddle-point action.

\section{Saddle Point Approximation for general Rashba-Dresselhaus spin-orbit coupling}
\label{sec:appendix-B}

We first analyze the saddle-point contribution. The saddle-point thermodynamic potential
$\Omega_0 = -T \ln {\cal Z}_0$ can be obtained for the saddle-point partition function
${\cal Z} = e^{-S_0}$  as $\Omega_0 = T S_0$.
Transforming the saddle-point Lagrangian ${\cal L}_0$ from
Eq.~(\ref{eqn:saddle-point-lagrangian-appendix}) into momentum space and integrating over spatial 
coordinates and imaginary time leads to the saddle-point thermodynamic potential,
\begin{equation}
\Omega_0 
= 
V\frac{|\Delta_0|^2}{g} 
- 
\frac{T}{2} \sum_{{\bf k},j} 
\ln (1 + e^{-\beta E_{{\bf k},j}}) 
+ 
\sum_{\bf k} \widetilde{K}_+ ({\bf k}),
\end{equation}
where $K_\alpha ({\bf k}) = {\bf k}^2 /2 m_\alpha - \mu_\alpha$ and 
the eigenvalues $E_{{\bf k},j}$ are the poles of ${\bf G}_0 ({\bf k},z)$, with
$j = \{1,2,3,4\}$.

Next, we restrict our analysis to mass 
balanced systems ($m_{\uparrow} = m_{\downarrow}$) in diffusive equilibrium 
($\mu_{\uparrow} = \mu_{\downarrow}$).  We also consider the general 
Rashba-Dresselhaus (GRD) spin-orbit field
$
h_\perp ({\bf k}) 
= \kappa (k_x + i \eta k_y) 
/ 
m,
$
where $\kappa$ and $\eta$ are the magnitude and anisotropy of the spin-orbit 
coupling, respectively.  Note that this form is equivalent to another common form of the 
Rashba-Dresselhaus coupling found in the literature~\cite{seo-2012a, seo-2012b}: 
${\bf h}_{so} = {\bf h}_R + {\bf h}_D$ 
where ${\bf h}_R = v_R (k_x \hat{{\bf y}} - k_y \hat{{\bf x}})$ 
and ${\bf h}_D = v_D (k_x \hat{{\bf y}} + k_y \hat{{\bf x}})$.  
The two forms are related via a momentum-space 
rotation and the correspondences 
$\kappa = m (v_R + v_D)$ and 
$\eta = (v_R - v_D) / (v_R + v_D)$.
The equal Rashba-Dresselhaus limit (ERD) corresponds to $v_R = v_D = v$, 
leading to $\eta = 0$ and $\kappa = 2 m v$.
The specific case of equal Rashba-Dresselhaus spin-orbit coupling
discussed in the main part of the paper corresponds to the case where $\eta = 0$,
that is,
$
h_\perp ({\bf k}) 
= \kappa k_x/ m.
$

For the general Rashba-Dresselhaus case, the four eigenvalues 
are 
\begin{eqnarray}
\label{eqn:eigenvalues-12-appendix}
E_{1,2} ({\bf k})
& = &
\left[
\zeta_{\bf k}^2 
\pm 2
\sqrt{
E_{0,{\bf k}}^2 h_{\bf k}^2
- |\Delta_0|^2 |h_\perp ({\bf k})|^2} \right]^{1/2},  \\
\label{eqn:eigenvalues-34-appendix}
E_{3,4} ({\bf k}) & = & - E_{2,1} ({\bf k}),
\end{eqnarray}
where the $+$ $(-)$ sign within the outermost square root
corresponds to $E_1$ $(E_2)$, and the functions inside the square roots are
$\zeta_{\bf k}^2 = E_{0, {\bf k}}^2 +  h_{\bf k}^2$, with contributions
\begin{eqnarray}
& & E_{0, {\bf k}}  =  \sqrt{\xi^2_{\bf k} + |\Delta_0|^2},   \\
& & h_{\bf k}  =  \sqrt{|h_\perp({\bf k})|^2 + h^2_z} ,
\end{eqnarray}
where
$
\xi_{\bf k} = \varepsilon_{\bf k} - \mu,
$ 
and 
$
\varepsilon_{\bf k} = {\bf k}^2 / 2 m .
$  
The order-parameter equation is found from the saddle point condition
$\delta \Omega_0 / \delta \Delta_0^* \vert_{T, V, \mu} = 0$. At the phase 
boundary between the superfluid and normal phases, $\Delta_0 \to 0$, and 
the order-parameter equation becomes 
\begin{eqnarray}
\label{eqn:order-parameter-appendix}
\frac{m}{4 \pi a_s} & = & \frac{1}{2 V}
\sum_{\bf k} \bigg[ \frac{1}{\varepsilon_{\bf k}}
- \frac{\tanh(\beta E_1/2)}{2 E_1} - \frac{\tanh(\beta E_2/2)}{2 E_2}   \nonumber   \\
& - & \frac{h^2_z}{\xi_{\bf k} h_{\bf k}} \hspace{.5mm}
\left(\frac{\tanh(\beta E_1/2)}{2 E_1}
- \frac{\tanh(\beta E_2/2)}{2 E_2} \right) \bigg],  \nonumber   \\ 
\end{eqnarray}
after expressing the interaction parameter $g$ in terms of the $s$-wave scattering
length via the relation
\begin{equation}
\label{eqn:scattering-length-appendix}
\frac{1}{g} =
-  \frac{m}{4 \pi a_s} + \frac{1}{V}\sum_{\bf k} \frac{1}{2 \varepsilon_{\bf k}}   .
\end{equation}
We note that $a_s$ is the $s$-wave scattering length 
\textit{in the absence} of spin-orbit and Zeeman fields.  
It is, of course, possible to express all relations obtained 
in terms of a scattering length which is renormalized by the presence 
of the spin-orbit and Rabi fields~\cite{goldbart-2011, ozawa-2012}.  
However, in addition to complicating our already cumbersome expressions, 
it would make reference to a quantity that is more difficult to measure 
experimentally and that would hide the explicit dependence of 
the properties that we analyze in terms of the 
spin-orbit and Rabi fields, so we do not consider such complications here.
Note that since $\Delta_0 = 0$ at the phase boundary, the eigenvalues
in Eq. (\ref{eqn:eigenvalues-12-appendix}) reduce
to $E_{1}({\bf k}) = \Big\vert \vert \xi_{\bf k}\vert + h_{\bf k} \Big\vert$,
$E_{2}({\bf k}) = \Big\vert \vert \xi_{\bf k}\vert - h_{\bf k} \Big\vert$,
which is the absolute value of the normal-state energy dispersions.
However, it is straightforward to show that
ignoring the absolute values does not result in any change in either the
mean-field order parameter given by Eq.~(\ref{eqn:order-parameter-appendix})
or number equation shown in Eq.~(\ref{eqn:number-saddle-point-appendix}),
when $\Delta_0 \to 0$.

The saddle-point critical temperature $T_0$ is determined by solving
Eq.~(\ref{eqn:order-parameter-appendix})
subject to the thermodynamic constraint 
$
N_0 = - \partial \Omega_0 / \partial \mu \vert_{T, V},
$ 
which yields
\begin{eqnarray}
\label{eqn:number-saddle-point-appendix}
  N_0 & = & \sum_{\bf k} \bigg \{ 1 - \xi_{\bf k} \bigg[ \frac{1}{\varepsilon_{\bf k}} + \frac{\tanh(\beta E_1/2)}{2 E_1} + \frac{\tanh(\beta E_2/2)}{2 E_2} \nonumber   \\
	&+ &  \frac{|h_\perp ({\bf k})|^2}{\xi_{\bf k} h_{\bf k}} \hspace{.5mm} \left(\frac{\tanh(\beta E_1/2)}{2 E_1} - \frac{\tanh(\beta E_2/2)}{2 E_2} \right) \bigg]  \bigg \}.   \nonumber   \\
\end{eqnarray}
%
A mean-field description of the system, which involves a simultaneous solution
of Eqs.~(\ref{eqn:order-parameter-appendix})
and~(\ref{eqn:number-saddle-point-appendix}),
yields the asymptotically correct description of the system in the BCS limit; however,
such a description fails miserably in the BEC regime where it does not account for the
formation of two-body bound states.
The general Rashba-Dresselhaus spin-orbit saddle-point equations~(\ref{eqn:order-parameter-appendix})
and~(\ref{eqn:number-saddle-point-appendix})
reduce to the equal Rashba-Dresselhaus equations~(\ref{eqn:saddle-point-order-parameter})
and~(\ref{eqn:saddle-point-number}) of the main part of the paper
with the explicit use of $h_z = \Omega_R/2$ and $h_\perp ({\bf k}) = \kappa k_x/m$,
where $\Omega_R$ is the Rabi coupling.

\section{Derivation of the modified number equation with Gaussian
fluctuations}
\label{sec:appendix-C}
We begin by deriving the modified number equation arising from Gaussian 
fluctuations of the order parameter near the superfluid phase boundary.  
The fluctuation thermodynamic potential $\Omega_F$ results from the Gaussian
integration of the fields $\eta ({\bf r}, \tau)$ and $\eta^{*} ({\bf r}, \tau)$
in the fluctuation partition function ${\cal Z}_{F} = \int d \eta^{*} d \eta e^{-S_F}$,
where the action $ S_{F} = \int d\tau_0^\beta \int d^3{\bf r} {\cal L}_F ({\bf r}, \tau)$ is calculated
to quadratic order. The contribution to the thermodynamic potential due to
Gaussian fluctuations is 
\begin{equation}
\Omega_F 
= 
- T\sum_{i q_n, {\bf q}} 
\ln 
\left[
\beta \Gamma ({\bf q},i q_n)/V
\right]
\end{equation}
where 
$q_n = 2 \pi n T$ are the bosonic Matsubara frequencies and 
$\Gamma ({\bf q},i q_n)$ is directly related to the pair fluctuation 
propagator 
$
\chi_{pair} ({\bf q}, i q_n) 
= 
V 
\Gamma^{-1} ({\bf q},i q_n).
$

The Matsubara sum can be evaluated via contour integration,
\begin{equation}
\label{eq:NSR}
\Omega_F 
= 
- 
T 
\sum_{\bf q} 
\oint_{\cal{C}} 
\frac{d z}{2 \pi i} 
\hspace{.5mm} 
n_B (z) 
\ln
\left[
\beta \Gamma ({\bf q}, z)/V
\right],   
\end{equation}
where $n_B (z) = 1 / (e^z - 1)$ 
is the Bose function and the countour ${\cal{C}}$ 
encloses all of the Matsubara poles of the Bose function.
Next, we deform the contour around the Matsubara frequencies towards 
infinity, taking into account the branch cut and the possibility of poles 
coming from the logarithmic term inside the countour integral.
We take the branch cut to be along the real axis, then add 
and subtract the pole at $i q_n = 0$ to obtain
\begin{equation}
\Omega_F 
= 
- 
T 
\sum_{\bf q} 
\int^\infty_{-\infty} 
\frac{d \omega}{\pi} 
\hspace{.5mm} 
n_B (\omega) 
\left[ 
\delta ({\bf q},\omega) 
- 
\delta ({\bf q},0) 
\right],
\end{equation}
where the phase shift $\delta ({\bf q}, \omega)$  is defined via 
$
\Gamma ({\bf q},\omega \pm i \epsilon) 
= 
|\Gamma ({\bf q},\omega)| 
e^{\pm i \delta ({\bf q},\omega)},
$ 
and arises from the contour segments above 
and below the real axis.
  
The thermodynamic identity 
$
N 
= 
- \partial \Omega / \partial \mu \vert_{T,V}
$ 
then yields to the fluctuation correction,
\begin{equation}
\label{eqn:number-fluctuation-appendix}
N_F 
= 
T
\sum_{\bf q} 
\int^\infty_{-\infty} 
\frac{d \omega}{\pi} 
\hspace{.5mm} 
n_B (\omega) 
\left[
\frac{\partial \delta ({\bf q},\omega)}{\partial \mu} 
- 
\frac{\partial \delta ({\bf q},0)}{\partial \mu} 
\right]   ,
\end{equation}
to the the saddle-point number equation, and has a similar 
analytical structure as in the case without spin-orbit and Zeeman
fields~\cite{nozieres-1985, sademelo-1993}. Thus, we can write the final
number equation at the critical temperature $T_c$ as $N = N_0 + N_F$.
Since the phase shift $\delta ({\bf q},z)$ vanishes everywhere that 
$\Gamma ({\bf q},z)$ is analytic, the only contributions to 
Eq.~(\ref{eqn:number-fluctuation-appendix}) arise from a possible isolated pole at 
$\omega_p ({\bf q})$ and a branch cut extending from
the two-particle continuum threshold 
$
\omega_{tp} ({\bf q}) 
= 
\min_{ \{ i,j,{\bf k} \} } 
\left[
E_{i} ({\bf k}) 
+ 
E_{j} ({\bf k+q})
\right]
$ 
to 
$z \to \infty$ along the positive real axis. The explicit
form of $\Gamma ({\bf q}, z)$ can be extracted 
from Eq.~(\ref{eqn:gamma-appendix}) of Appendix~\ref{sec:appendix-D}.

When there is a pole corresponding to the emergence of a two-body bound state,
we can explicitly write
$
\Gamma ({\bf q},z) 
\sim R ({\bf q}) 
/ 
(z - \omega_p ({\bf q})),
$
from which we obtain 
$
\partial \delta ({\bf q},\omega) 
/ 
\partial \mu 
= 
2 \delta (z - \omega_p ({\bf q})),
$ 
leading to the bound state density
\begin{equation}
\label{eqn:number-bound-states-appendix}
  N_ b 
= 
2 \sum_{\bf q} n_B (\omega_p ({\bf q})),
\end{equation}
where the energy $\omega_{p} ({\bf q})$ must lie below the two-particle
continuum threshold $\omega_{tp} ({\bf q})$.
The factor of 2, which arises naturally, is due to the two fermions 
comprising a bosonic molecule.  Naturally, the presence of this term 
in the fluctuation-modified number equation is dependent upon the 
existence of such a pole, that is, a molecular bound state. These bound
states correspond to the Feshbach molecules in the presence of
spin-orbit coupling and Zeeman fields~\cite{spielman-2013a, doga-2016}.

Having extracted the pole contribution 
to Eq.~(\ref{eqn:number-fluctuation-appendix}), when it exists,
the remaining integral over the branch cut corresponds to scattering state 
fermions,
\begin{equation}
\label{eqn:number-scattering-appendix}
N_{sc} 
= 
T 
\sum_{\bf q} 
\int^\infty_{\omega_{tp} ({\bf q})} 
\frac{d \omega}{\pi} \hspace{.5mm} 
n_B (\omega) 
\left[ 
\frac{\partial \delta ({\bf q},\omega)}{\partial \mu} 
- 
\frac{\partial \delta ({\bf q},0)}{\partial \mu} 
\right],   
\end{equation}
whose energy is larger than the minimum energy 
$\omega_{tp} ({\bf q})$ of two free fermions.
Thus, when bound states are present, we arrive at the modified number equation,
\begin{equation}
N
= 
N_0 + N_{sc} + N_b
\end{equation}
where $N_0$ is the number of free fermions obtained from the 
saddle-point analysis in Eq.~(\ref{eqn:number-saddle-point-appendix}),
and $N_b$ and $N_{sc}$ are the bound state and scattering contributions given
in Eqs.~(\ref{eqn:number-bound-states-appendix})
and~(\ref{eqn:number-scattering-appendix}),
respectively. These general results are particularized to the equal Rashba-Dresselhaus case in
Sec.~\ref{sec:gaussian-fluctuations} of this paper.
   
The number of unbound states $N_{u}$ is then easily seen to be
$N_u = N_0 + N_{sc}$, that is, the sum of the free-fermion $(N_0)$ and
scattering $(N_{sc})$ contributions.
Naturally, the number of unbound states is also equal to the total number
of states, $N$, minus the number of bound states, $N_b$, that is, $N_u = N - N_b$.

\section{Derivation of Ginzburg-Landau coefficients for
  general Rashba-Dresselhaus spin-orbit coupling}
\label{sec:appendix-D}

Next, we derive explicit expressions for the coefficients 
of the time-dependent Ginzburg-Landau theory valid near the critical
temperature of the superfluid. We start from the fluctuation Lagrangian,
\begin{equation}
{\cal L}_F ({\bf r}, \tau) =  
- \frac{T}{2 V} 
\hspace{.5mm} 
{\rm Tr} \ln ({\bf I} + {\bf G}_0 {\bf G}^{-1}_F) 
+
\Lambda ({\bf r}, \tau) 
+
\frac{|\eta ({\bf r},\tau)|^2}{g},   
\label{eq:L_fluc2}
\end{equation}
in a volume $V$, and take the traces over 
both discrete and continuous indices. Notice that the term 
$
\Lambda ({\bf r}, \tau) 
= 
\left[
\Delta_0 \eta^* ({\bf r},\tau) 
+ 
\Delta_0^* \eta ({\bf r},\tau) 
\right]/g
$
in the fluctuation Lagrangian cancels out the linear terms in 
$\eta$ and $\eta^*$ when the logarithm is expanded, due to the saddle-point 
condition.  Since the expansion is performed near $T_c$, 
we take the saddle-point order parameter $\Delta_0 \to 0$ and redefine the
fluctuation field as
$
\eta ({\bf r}, \tau) 
= 
\Delta ({\bf r}, \tau)
$
to obtain 
\begin{equation}
{\cal L}_F ({\bf r}, \tau)
= 
\frac{|\Delta|^2}{g} 
- 
\frac{T}{2 V} 
{\rm Tr} 
\ln ({\bf I} + {\bf G}_0 [0] {\bf G}^{-1}_F [\Delta] ).
\end{equation}
Notice that the arguments in 
${\bf G}_0 [0]$ and ${\bf G}^{-1}_F [\Delta]$ 
represent the values of $\Delta_0 = 0$ and $\eta = \Delta$, respectively.  

We expand the logarithm to sixth order in $\Delta$ to obtain
\begin{eqnarray}
\label{eqn:fluctuation-lagrangian-appendix}
{\cal L}_F ({\bf r}, \tau)
& = &
\frac{|\Delta|^2}{g} 
+ 
\frac{T}{2 V} 
{\rm Tr} 
\bigg[ 
\frac{1}{2}
({\bf G}_0 {\bf G}^{-1}_F)^2 
+ 
\frac{1}{4} ({\bf G}_0 {\bf G}^{-1}_F)^4 
\nonumber
\\
& & + \frac{1}{6} ({\bf G}_0 {\bf G}^{-1}_F)^6 + ... 
\bigg],   
\end{eqnarray}
where the higher-order odd (cubic and quintic) terms in the order-parameter 
amplitudes expansion can be shown to vanish due to conservation laws
and energy or momentum considerations.

The traces can be evaluated explicitly 
by using the momentum-space inverse single-particle Green's function
\begin{equation}
\label{eqn:green-zero-inverse-appendix}
{\bf G}^{-1}_0 (k,k^\prime) 
= 
\begin{pmatrix}
{\bf A}^{-1} (k) & {\bf 0} \\
{\bf 0} & - \left[ {\bf A}^{-1} (-k) \right]^T
\end{pmatrix} 
\delta_{k k^\prime},   
\end{equation}
derived from Eq.~(\ref{eqn:greens-function-appendix}).
Here, we use the shorthand notation $k \equiv (i\omega, {\bf k})$,
where $\omega_n = 2\pi n T$ are bosonic Matsubara frequencies 
and define the $2 \times 2$ matrix,
\begin{equation}
{\bf A}^{-1} (k) 
= 
\begin{pmatrix}
i\omega_n - \widetilde{K}_{\uparrow} ({\bf k}) & h^*_\perp ({\bf k})   \\
h_\perp ({\bf k}) & i\omega_n - \widetilde{K}_{\downarrow} ({\bf k})
\end{pmatrix},   
\label{eq:Ainv}
\end{equation}
where $\widetilde{K}_{\uparrow} = \xi_\mathbf{k} - h_z$,
$\widetilde{K}_{\downarrow} = \xi_\mathbf{k} + h_z$, with 
$\xi_\mathbf{k} = \mathbf{k}^2 / 2m - \mu$ the kinetic energy
relative to the chemical potential, $h_z$ the external Zeeman
field, and $h_\perp ({\bf k}) = h_x ({\bf k})  + i h_y ({\bf k})$ the 
spin-orbit field.  We also define the fluctuation contribution 
to the inverse Green's function,
\begin{equation}
{\bf G}^{-1}_F (k, k^\prime) 
= 
\begin{pmatrix}
{\bf 0} & -i \bm{\sigma}_y \Delta_{k - k^\prime}   \\
i \bm{\sigma}_y \Delta^\dagger_{k^\prime - k} & {\bf 0}
\end{pmatrix},
\end{equation}
where $\bm{\sigma}_y$ is the second Pauli matrix in isospin (internal state) space 
and 
\begin{equation}
\Delta_k 
= 
\frac{\beta}{V}
\int_0^\beta  d\tau \int d^3 {\bf r} 
e^{i ({\bf k} \cdot {\bf r} - \omega \tau)} 
\Delta (r) 
\end{equation}
is the Fourier transform of $\Delta (r)$, with $r \equiv ({\bf r}, \tau)$, and also has dimensions
of energy. Recall that we set $\hbar = k_B = 1$, 
such that energy, frequency and temperature have the same units.

Inversion of Eq.~(\ref{eqn:green-zero-inverse-appendix}) yields
\begin{equation}
{\bf G}_0 (k,k^\prime) 
= 
\begin{pmatrix}				
{\bf A} (k) & {\bf 0} \\
{\bf 0} & - \left[ {\bf A} (-k) \right]^T
\end{pmatrix} 
\delta_{k k^\prime},   
\label{eq:G0}
\end{equation}
where the matrix ${\bf A} (k)$ is
\begin{equation}
{\bf A} (k) = 	\frac{1}{\det [{\bf A}^{-1} (k)]}
\begin{pmatrix}
i\omega_n - \widetilde{K}_{\downarrow} ({\bf k}) & -h^*_\perp ({\bf k})   \\
-h_\perp ({\bf k}) & i\omega_n - \widetilde{K}_\uparrow {\bf k})
\end{pmatrix}.   
\label{eq:A}
\end{equation}
with 
$
\det [{\bf A}^{-1} (k)] 
= 
\prod^2_{j=1}
\left[
i\omega_n - E_{j} ({\bf k})
\right]
$
and where the independent-particle eigenvalues $E_{j} ({\bf k})$ are 
two of the poles of ${\bf G}_0 (k,k)$. These poles are exactly the general
eigenvalues described in Eqs.~(\ref{eqn:eigenvalues-12-appendix})
in the limit of $\Delta_0 \to 0$.
Note that setting $\Delta_0 = 0$ in the general eigenvalue 
expressions yields 
$E_{1,2} ({\bf k}) = \left| |\xi_{\bf k}| \pm h_{\bf k} \right|$.
The other set of poles of ${\bf G}_0 (k,k)$ corresponds to the eigenvalues
$E_{3,4} ({\bf k}) = - E_{2, 1} ({\bf k})$ found 
from $\det \left[ {\bf A}^{-1} (-k) \right]^T = 0$.

Using Eq.~(\ref{eqn:fluctuation-lagrangian-appendix}) to write the fluctuation action as
$
{\cal S}_F 
= 
\int_0^\beta d\tau \int d^3{\bf r}{\cal L}_F ({\bf r}, \tau),
$
results in
\begin{eqnarray}
\label{eqn:ginzburg-landau-fluctuation-action-appendix}
{\cal S}_F
& = & 
\beta V \sum_q \frac{|\Delta_q|^2}{\Gamma (q)} 
+ 
\frac{\beta V }{2} \sum_{q_1,q_2,q_3} 
b_{1,2,3} 
\Delta_{1} \Delta^*_{2} \Delta_{3} \Delta^*_{1 - 2 + 3}   
\hspace{5mm} \nonumber   \\
&& \hspace{0mm} 
+ 
\frac{\beta V}{3} \sum_{q_1 \cdots q_5}
f_{1 \cdots 5} 
\Delta_{1} \Delta^*_{2} \Delta_{3} 
\Delta^*_{4} \Delta_{5} \Delta^*_{1 - 2 + 3 -  4 + 5},
\end{eqnarray}
where summation over $q \equiv (i q_n, {\bf q})$ indicates sums over 
both the bosonic Matsubara frequencies $q_n = 2 \pi n T$ 
and momentum ${\bf q}$. Here, we used the shorthand notation
$j \equiv q_j$ to represent the labels of $\Delta_{q_j}$ or  $\Delta_{q_j}^*$.

The quadratic order appearing 
in Eq.~(\ref{eqn:ginzburg-landau-fluctuation-action-appendix}) 
arises from the terms $\vert \Delta ({\bf r}, \tau)\vert^2/g$ and  
$(T/2V) {\rm Tr} ({\bf G}_0 {\bf G}^{-1}_F)^2/2$ 
in Eq.~(\ref{eqn:fluctuation-lagrangian-appendix}), and is directly
related to the pair propagator $\chi_{pair} (q) = V \Gamma^{-1} (q)$,
with
\begin{equation}
\Gamma^{-1} (q) 
= 
\frac{1}{g} - 
\frac{T}{2 V} 
\sum_k 
\frac{{\rm Tr}\left[ {\bf A} (k) {\bf A}^{-1} (q - k) \right]}
{\det [{\bf A}^{-1} (q - k)]},   
\label{eq:sm-gamma-neg-one}
\end{equation}
where we use the identity 
$
\bm{\sigma}_y {\bf A} \bm{\sigma}_y 
= 
{\rm det} ({\bf A}) ({\bf A}^T)^{-1}.
$ 
%
%
\begin{widetext}
The fourth-order contribution arises from 
$\frac{1}{4} ({\bf G}_0 {\bf G}^{-1}_F)^4$ and
leads to 
\begin{equation}
b (q_1, q_2, q_3) 
= 
\frac{T}{2 V} 
\sum_k 
\frac{
{\rm Tr} 
\left[ 
{\bf A} (k) {\bf A}^{-1} (q_1 - k) 
{\bf A} (k - q_1 + q_2) {\bf A}^{-1} (q_1 - q_2 + q_3 - k) 
\right]
}
{
{\rm det} 
\left[
{\bf A}^{-1} (q_1 - k)
\right] 
{\rm det} 
\left[
{\bf A}^{-1} (q_1 - q_2 + q_3 - k)
\right]
},
\label{eq:sm-b}
\end{equation}
while the sixth order contribution emergences from 
$\frac{1}{6} ({\bf G}_0 {\bf G}^{-1}_F)^6$, giving
\begin{eqnarray}
f (q_1, \cdots , q_5) 
& = & 
\frac{T}{2 V} 
\sum_k {\rm det} 
\left[
{\bf A} (q_1 - k)
\right] 
{\rm det} 
\left[
{\bf A} (q_1 - q_2 + q_3 - k)
\right] 
{\rm det} 
\left[
{\bf A} (q_1 - q_2 + q_3 - q_4 + q_5 - k)  
\right]   
\nonumber   \\
&& 
\hspace{15mm} 
\times {\rm Tr} 
\bigg[ 
{\bf A} (k) {\bf A}^{-1} (q_1 - k) 
{\bf A} (k - q_1 + q_2) {\bf A}^{-1} (q_1 - q_2 + q_3 - k)  
 \nonumber   \\
&& 
\hspace{25mm} 
\times {\bf A} (k - q_1 + q_2 - q_3 + q_4) 
{\bf A}^{-1} (q_1 - q_2 + q_3 - q_4 + q_5 - k) 
\bigg].
\label{eq:sm-e}
\end{eqnarray}
\end{widetext}

Evaluating the expressions given in Eqs.~(\ref{eq:sm-gamma-neg-one})
through~(\ref{eq:sm-e}) requires us to perform summations 
over Matsubara frequencies of the type
\begin{equation}
T \sum_{i \omega_n} \frac{1}{i \omega_n \pm E ({\bf k})} 
= 
\begin{cases}
\hspace{3mm} n ({\bf k})   \hspace{6mm} \mbox{if ``$+$''}   \\
1 - n ({\bf k})  \hspace{3mm} \mbox{if ``$-$''}   \\
\end{cases},
\end{equation}
where $n ({\bf k}) = 1 / \left[ e^{\beta E ({\bf k})} + 1 \right]$ is the Fermi function.
For the quadratic term, we obtain the result
\begin{eqnarray}
\label{eqn:gamma-appendix}
\Gamma^{-1} ({\bf q}, i q_n)
& = &
- \frac{m}{4 \pi a_s} 
+ \frac{1}{2 V} \sum_{\bf k} 
\bigg[
\frac{1}{\varepsilon_{\bf k}}
\nonumber \\
& + & 
\sum_{i, j = 1}^2  
\alpha_{ij} ({\bf k}, {\bf q}) 
W_{ij} ({\bf k}, {\bf q}, iq_n)
\bigg],
\end{eqnarray}
where the functions in the last term are 
\begin{equation}
W_{ij} ({\bf k}, {\bf q}, iq_n)
= 
\frac{1 - n_i ({{\bf k}}) - n_{j}({\bf k}+{\bf q})}{i q_n - E_{i} ({\bf k}) - E_{j} ({\bf k}+{\bf q})},
\end{equation}
corresponding to the contribution of bubble diagrams
to the pair susceptibility. The coherence factors are
\begin{eqnarray}
\alpha_{11} ({\bf k}, {\bf q})
& = & 
\vert 
u_{\bf k} u_{{\bf k}+{\bf q}} 
- 
v_{\bf k} v^*_{{\bf k}+ {\bf q}}
\vert^2,
\\
\alpha_{12} ({\bf k}, {\bf q})
& = &
\vert
u_{\bf k} v_{{\bf k}+{\bf q}} + u_{{\bf k}+{\bf q}} v_{\bf k}
\vert^2,
\end{eqnarray}
with
$
\alpha_{11} ({\bf k}, {\bf q})
=
\alpha_{22} ({\bf k}, {\bf q})
$
and
$
\alpha_{12} ({\bf k}, {\bf q})
=  
\alpha_{21} ({\bf k}, {\bf q}),
$
where the quasiparticle amplitudes are
\begin{eqnarray}
\label{eqn:amplitudes-u}  
u_{\bf k} 
= 
\sqrt{\frac{1}{2} 
\left(
1 + \frac{h_z}{h_{\bf k}} 
\right)} ,  
\\
\label{eqn:amplitudes-v}
v_{\bf k} 
= 
e^{i \theta_{\bf k}} 
\sqrt{\frac{1}{2} 
\left(
1 - \frac{h_z}{h_{\bf k}} 
\right)
}.
\end{eqnarray}
The angle $\theta_{\bf k}$ is the phase associated with the spin-orbit field 
$
h_\perp ({\bf k}) 
= 
\vert h_\perp ({\bf k}) \vert 
e^{i \theta_{\bf k}},
$ 
and we replaced the interaction parameter 
$g$ by the $s$-wave scattering length $a_s$
via Eq.~(\ref{eqn:scattering-length-appendix}),
recalling that $\varepsilon_{\bf k} = {\bf k}^2/2m$.
The phase and modulus of $h_{\perp} ({\bf k})$ are 
\begin{eqnarray}
\theta_{\bf k}
=
\arctan \left( \frac{\eta k_y}{k_x} \right),
 \\
\vert h_\perp ({\bf k}) \vert
=
\frac{\vert \kappa \vert}{m}
\sqrt{k_x^2 + \eta k_y^2},
\\
\nonumber
\end{eqnarray}
and the total effective field is
\begin{equation}
h_{\bf k} = \sqrt{h_z^2 + \vert h_{\perp} ({\bf k}) \vert^2}.
\end{equation}

Since we are interested only in the long-wavelength and low-frequency 
regime, we perform an analytic continuation to real frequencies 
$
iq_n = \omega + i\delta
$
after calculating the Matsubara sums for all coefficients
appearing in Eq.~(\ref{eqn:ginzburg-landau-fluctuation-action-appendix}) and perform 
a small momentum ${\bf q}$ and low-frequency $\omega$ expansion resulting
in the Ginzburg-Landau action,
\begin{widetext}
\begin{eqnarray}
{\cal S}_F 
= 
{\cal S}_{GL} 
& = & 
\beta V
\sum_{q}
\left( a + \sum_{\ell} c_\ell \frac{ q^2_\ell}{2 m} - d_0 \omega \right) 
|\Delta_q|^2 
+ 
\frac{\beta V}{2} 
\sum_{q_1, q_2, q_3}  b (q_1,q_2,q_3) 
\Delta_{q_1} \Delta^*_{q_2} \Delta_{q_3} \Delta^*_{q_1 - q_2 + q_3}   
\hspace{5mm} \nonumber   \\
&& \hspace{0mm} 
+ 
\frac{\beta V}{3} 
\sum_{q_1 \cdots q_5}
f (q_1, q_2, q_3, q_4, q_5) 
\Delta_{q_1} \Delta^*_{q_2} \Delta_{q_3} 
\Delta^*_{q_4} \Delta_{q_5} \Delta^*_{q_1 - q_2 + q_3 - q_4 + q_5}.
\label{eq:sm-fluctuation-action-GL}
\end{eqnarray}
\end{widetext}
Here, the label $\ell$ appearing explicitly in the term $\sum_{\ell} c_{\ell} q_{\ell}^2/(2m)$
represents the spatial directions $\{x, y, z \}$, while the $q_{j}$'s in the sums
correspond to $({\bf q}_{j}, \omega_{j})$ and 
the summations $\sum_{q_j}$ represent integrals 
$\beta V \int d\omega_j \int d^3 {\bf q}_{j}$, where $j$ labels a fermion
pair and can take values in the set $\{1, 2, 3, 4, 5\}$.
In the expression above, we used the result
\begin{equation}
\Gamma^{-1} ({\bf q},\omega) 
= 
a 
+ 
\sum_{\ell} c_{\ell} \frac{q_{\ell}^2}{2 m} 
- d_0 \omega 
+ 
\cdots
\label{eq:sm-gamma-neg-one-expansion}
\end{equation}
for the analytically continued expression of 
$\Gamma^{-1} ({\bf q},iq_n)$ appearing in Eq.~(\ref{eqn:gamma-appendix}). 
To write the coefficients above in a more compact notation, we define 
\begin{eqnarray}
X_i = X_i ({\bf k})
= 
\tanh \left[\beta E_i ({\bf k})/2\right],
\\
Y_i = Y_i ({\bf k}) 
= 
{\rm sech}^2 \left[\beta E_i ({\bf k})/ 2 \right].
\end{eqnarray}
The frequency- and momentum-independent coefficient is
\begin{eqnarray}
\label{eq:sm-a}
a  =
-  \frac{m}{4 \pi a_s} 
+ 
\frac{1}{V} 
\sum_{\bf k}
\bigg[ 
\frac{1}{2\varepsilon_{\bf k}}
 - 
\left( 
\frac{X_1}{4 E_1} 
+ 
\frac{X_2}{4 E_2} 
\right)
\nonumber
\\
 -  
\frac{h^2_z}{\xi_{\bf k} h_{\bf k}} 
\left( 
\frac{X_1}{4 E_1} - \frac{X_2}{4 E_2} 
\right)
\bigg],
\end{eqnarray}
where $E_1 = E_1 ({\bf k})$ and $E_2 = E_2 ({\bf k})$.
The coefficient $d_0 = d_R + i d_I$ multiplying the linear term 
in frequency has a real component given by 
\begin{equation}
d_R 
= 
\frac{1}{2 V} 
{\cal{P}} 
\sum_\mathbf{k} 
\sum^2_{i,j = 1} 
\alpha_{ij} (\mathbf{k},\mathbf{0}) 
\frac{1 - n_{i} ({\bf k}) - n_{j} ( {\bf k})}
 {
\left[
E_{i} ({\bf k}) + E_{j} ({\bf k})
\right]^2}.
\end{equation}
Using the explicit forms 
of the coherence factors $u_{\bf k}$ and $v_{\bf k}$ that define 
$\alpha_{ij} ({\bf k}, {\bf q} = {\bf 0})$, 
the above expression can be rewritten as
\begin{eqnarray}
d_R 
=  
\frac{1}{2 V} \hspace{.5mm} 
{\cal{P}} \sum_{\bf k} 
\biggr[ 
\left( 1 + \frac{h^2_z}{\xi^2_{\bf k}} 
\right) 
\left( 
\frac{X_1}{4 E^2_1} + \frac{X_2}{4 E^2_2} 
\right)
\nonumber
\\
+  
\frac{2 h^2_z}{\xi_{\bf k} h_{\bf k}} 
\left( \frac{X_1}{4 E^2_1} - \frac{X_2}{4 E^2_2} 
\right) 
\biggr],
\label{eq:sm-dR}   
\end{eqnarray}
which defines the time scale for temporal oscillations of the order
parameter. Here, the symbol ${\cal{P}}$ denotes the principal value,
and the coefficient $d_R$ is obtained from
\begin{widetext}
\begin{equation}
{\rm Re}
\left[
 \Gamma^{-1} ({\bf q} = {\bf 0}, \omega + i\delta)
\right]
= 
- \frac{m}{4 \pi a_s} 
+ \frac{1}{2 V} \sum_{\bf k} 
\left[ 
\frac{1}{\varepsilon_{\bf k}} 
+ 
{\cal P} \sum^2_{i,j = 1} 
\alpha_{ij} ({\bf k}, {\bf q} = {\bf 0}) 
\frac{1 - n_{i} ({\bf k})- n_{j} ({\bf k})}{ \omega - E_{i} ({\bf k})  - E_{j} ({\bf k}) }
\right].
\end{equation}

The imaginary component of the coefficient $d$ has the form
\begin{equation}
d_I 
= 
\frac{\pi}{2 V} 
\sum_\mathbf{k} \sum^2_{i,j = 1} 
\alpha_{ij} (\mathbf{k},\mathbf{0}) 
\left[ 1 - n_{i} ({\bf k}) - n_{j} ({\bf k}) \right] 
\delta^\prime  \left( E_{i} ({\bf k}) + E_{j} ({\bf k}) \right),
\end{equation}
where the derivative of the delta function is
$
\delta^\prime (\lambda) 
= 
\partial \delta (x + \lambda) 
/
\partial x \vert_{x = 0}.
$
Using again the expressions of the coherence factors $u_{\bf k}$ and $v_{\bf k}$ 
leads to 
\begin{equation}
d_I 
=  
\frac{\pi}{2 V} 
\sum_{\bf k} 
\left \{ 
(X_1 + X_2) \delta^\prime (2 \xi_{\bf k}) 
+ 
\frac{|h_\perp|^2}{h^2_{\bf k}} 
\bigg[ 
X_1 \delta^\prime (2 E_1) 
+ 
X_2 \delta^\prime (2 E_2) 
- 
(X_1 + X_2) 
\delta^\prime (2 \xi_{\bf k}) 
\bigg] 
\right \}, 
\label{eq:sm-dI}   
\end{equation}
which determines the lifetime of fermion pairs. This result originates from 
\begin{equation}
{\rm Im} 
\left[
\Gamma^{-1} ({\bf q} = {\bf 0}, \omega + i \delta)
\right]
= 
-\frac{\pi}{2V} 
\sum_{\bf k}
\sum_{i, j = 1}^{2}
\alpha_{ij} ({\bf k},{\bf q} = {\bf 0}) 
\left[
1 - n_{i} ({\bf k}) - n_{ j} ({\bf k}) 
\right]
\delta \left( \omega - E_{i} ({\bf k}) - E_ j ({\bf k}) \right),
\end{equation}
\end{widetext}
which immediately reveals that below the two-particle threshold
$
\omega_{tp} ({\bf q} = {\bf 0}) 
= 
{\rm min}_{\{i,j, {\bf k} \}} 
\left[
E_{i} ({\bf k}) 
+
E_{j} ({\bf k})
\right]
$ 
at center-of-mass momentum ${\bf q} = {\bf 0}$, the lifetime of the pairs is 
infinitely long due to the emergence of stable two-body bound states.
Note that collisions between bound states are not yet included.

The expressions for the $c_{\ell}$ coefficients appearing in 
Eq.~(\ref{eq:sm-gamma-neg-one-expansion}) are quite long and complex.
Since these coefficients are responsible for the mass renormalization and anisotropy
within the Ginzburg-Landau theory, we outline below their derivation in detail.
These coefficients can be obtained from the last term in
Eq.~(\ref{eqn:gamma-appendix}), which we define as
\begin{equation}
\label{eqn:function-f}
F ({\bf q}) =
\frac{1}{2 V} \sum_{\bf k} 
\sum^2_{i,j = 1} 
\alpha_{ij} ({\bf k}, {\bf q}) 
W_{ij} ({\bf k}, {\bf q}, iq_n = 0).
\end{equation}
The relation between $c_{\ell}$ and the function $F ({\bf q})$ defined above is 
\begin{equation}
\label{eqn:c-ell-coefficients}
c_{\ell} =
m \left[ \frac{\partial^2 F ({\bf q})}{\partial q_{\ell}^2} \right]_{{\bf q} = {\bf 0}}.
\end{equation}
A more explicit form of $c_{\ell}$ 
is obtained by analyzing the symmetry properties of $F({\bf q})$ under inversion
and reflection symmetries. To make these properties clear, we rewrite the summand
in Eq.~(\ref{eqn:function-f}) by making use of the transformation
${\bf k} \to {\bf k} - {\bf q}/2$. This procedure leads to the symmetric form,
\begin{equation}
\label{eqn:function-f-symmetric}
F ({\bf q}) =
\frac{1}{2V} \sum_{\bf k} \sum_{i,j = 1}^2
{\widetilde \alpha}_{ij} ({\bf k}_{-}, {\bf k}_{+})
{\widetilde W}_{ij} \left[ E_i ({\bf k}_{-}), E_j ({\bf k}_{+}) \right].
\end{equation}
Here, ${\bf k}_{+} = {\bf k} + {\bf q}/2$ and ${\bf k}_{-} = {\bf k} - {\bf q}/2$ are
new momentum labels, and 
\begin{eqnarray}
\label{eqn:alpha-11}
{\widetilde \alpha}_{11} ({\bf k}_{-}, {\bf k}_{+})
& = &
\vert u_{{\bf k}_{-}} u_{{\bf k}_{+}} - v_{{\bf k}_{-}} v^*_{{\bf k}_{+}} \vert^2 ,
\\
\label{eqn:alpha-12}
{\widetilde \alpha}_{12} ({\bf k}_{-}, {\bf k}_{+})
& = & 
 \vert u_{{\bf k}_{-}} v_{{\bf k}_{+}} - v_{{\bf k}_{-}} u_{{\bf k}_{+}} \vert^2  
\end{eqnarray}
are coherence factors,
with
$
{\widetilde \alpha}_{11} ({\bf k}_{-}, {\bf k}_{+})
= 
 {\widetilde \alpha}_{22} ({\bf k}_{-}, {\bf k}_{+})
$
 and
$
{\widetilde \alpha}_{12} ({\bf k}_{-}, {\bf k}_{+})
= 
{\widetilde \alpha}_{21} ({\bf k}_{-}, {\bf k}_{+})
$
The functions $u_{{\bf k}_{\pm}}$ and $v_{{\bf k}_{\pm}}$ are defined in
Eqs.~(\ref{eqn:amplitudes-u}) and (\ref{eqn:amplitudes-v}).  It is now very easy to show that
${\widetilde \alpha}_{ij} ({\bf k}_{-}, {\bf k}_{+}) = {\widetilde \alpha}_{ij} ({\bf k}_{+}, {\bf k}_{-})$, that is,
${\widetilde \alpha}_{ij} ({\bf k}_{-}, {\bf k}_{+})$ is an even function of ${\bf q}$, since taking
${\bf q} \to - {\bf q}$ leads to ${\bf k}_{-} \to {\bf k}_{+}$ and  ${\bf k}_{+} \to {\bf k}_{-}$
leaving ${\widetilde \alpha}_{ij}$ invariant.
It is also clear, from its definition, that ${\widetilde \alpha}_{ij}$ is symmetric in the band indices $\{i, j \}$. 
Furthermore, the function
\begin{equation}
{\widetilde W}_{ij} \left[E_{i} ({\bf k}_{-}), E_{j} ({\bf k}_{+}) \right]
=
\frac{{\cal N}_{ij}}{{\cal D}_{ij}}, 
\end{equation}
defined above, is the ratio between the numerator,
\begin{equation}
{\cal N}_{ij} =
\tanh \left[ \beta E_i ({\bf k}_{-})/2\right]
+
\tanh \left[ \beta E_j ({\bf k}_{+})/2\right],
\end{equation}
representing the Fermi occupations and the denominator,
\begin{equation}
{\cal D}_{ij}
=
2 \left[ E_i ({\bf k}_{-}) + E_j ({\bf k}_{+}) \right],
\end{equation}
representing the sum of the quasi-particle excitation energies.
To elliminate the Fermi distributions $n_i ({\bf k})$ in the numerator,
we used the relation $1 - 2 n_i ({\bf k}) = \tanh \left[ \beta E_i ({\bf k}_{-})/2\right]$.
Notice that
${\widetilde W}_{ij} \left[E_{i} ({\bf k}_{-}), E_{j} ({\bf k}_{+}) \right]$
is not generally symmetric under inversion ${\bf q} \to - {\bf q}$, that is,
under the transformation ${\bf k}_{-} \to {\bf k}_{+}$ and
${\bf k}_{+} \to {\bf k}_{-}$. This means that
${\widetilde W}_{ij} \left[E_{i} ({\bf k}_{-}), E_{j} ({\bf k}_{+}) \right] \ne
{\widetilde W}_{ij} \left[E_{i} ({\bf k}_{+}), E_{j} ({\bf k}_{-}) \right] $,
unless when $i = j$, where it is trivially an even function of ${\bf q}$. However,
${\widetilde W}_{ij} \left[E_{i} ({\bf k}_{-}), E_{j} ({\bf k}_{+}) \right]$ is
always symmetric under simultaneous
momentum inversion $({\bf q} \to -{\bf q})$ and band index exchange, that is,
\begin{equation}
{\widetilde W}_{ij} \left[E_{i} ({\bf k}_{-}), E_{j} ({\bf k}_{+}) \right]
=
{\widetilde W}_{ji} \left[E_{j} ({\bf k}_{+}), E_{i} ({\bf k}_{-}) \right]
\end{equation}
for any $\{i, j \}$.
This property will be used later to write a final expression for $c_{\ell}$.
Next, we write
\begin{equation}
  \left[ \frac{\partial^2 F ({\bf q})}{\partial q_{\ell}^2} \right]_{{\bf q} = {\bf 0}}
  =
  \frac{1}{2V}
\sum_{\bf k} \sum_{i,j = 1}^2
{\cal F}_{ij},
\end{equation}
where the function inside the summation is 
\begin{equation}
{\cal F}_{ij}
=
\left[
\frac{\partial^2 {\widetilde \alpha}_{ij}}{\partial q_{\ell}^2} {\widetilde W}_{ij}
+
 \alpha_{ij} \frac{\partial^2 {\widetilde W}_{ij}}{\partial q_{\ell}^2}
\right]_{{\bf q} = {\bf 0}}.
\end{equation}
Notice the absence of terms containing the product of the first-order
derivatives of ${\widetilde \alpha}_{ij}$ and ${\widetilde W}_{ij}$.
These terms vanish due to parity since ${\widetilde \alpha}_{ij}$ is an even
function of ${\bf q}$, leading to
$\left[\partial {\widetilde \alpha}_{ij}/\partial q_{\ell} \right]_{{\bf q} = {\bf 0}} = 0$.
The last expression can be further developed upon summation
over the band indices, leading to
\begin{equation}
\left[ \frac{\partial^2 F ({\bf q})}{\partial q_{\ell}^2} \right]_{{\bf q} = {\bf 0}}
=
{\cal A} 
+
{\cal B}.
\end{equation}
The first contribution is given by 
\begin{equation}
{\cal A} 
=
\frac{1}{2V}
\sum_{{\bf k}}
\left[
\frac{\partial^2 {\widetilde \alpha}_{11}}{\partial q_{\ell}^2}
{\widetilde W}_{\rm di} 
+
\frac{\partial^2 {\widetilde \alpha}_{12}}{\partial q_{\ell}^2}
{\widetilde W}_{\rm od}
\right]_{{\bf q} = {\bf 0}},
\end{equation}
and contains the second derivatives of ${\widetilde \alpha}_{ij}$ and
the symmetric terms
\begin{eqnarray}
{\widetilde W}_{\rm di} = \left( {\widetilde W}_{11} + {\widetilde W}_{22} \right),
\\
{\widetilde W}_{\rm od} = \left( {\widetilde W}_{12} + {\widetilde W}_{21} \right),
\end{eqnarray}
The second contribution is given by 
\begin{equation}
{\cal B} 
=
\frac{1}{2V}
\sum_{{\bf k}}
\left[
  {\widetilde \alpha}_{11}
\frac{\partial^2 {\widetilde W}_{\rm di}}{\partial q_{\ell}^2} 
+
{\widetilde \alpha}_{12}
\frac{\partial^2 {\widetilde W}_{\rm od}}{\partial q_{\ell}^2} 
\right]_{{\bf q} = {\bf 0}}.
\end{equation}

Next, we explicitly write ${\widetilde \alpha}_{ij}$,
${\widetilde W}_{ij}$ and their second derivatives with respect to
$q_{\ell}$ at ${\bf q} = {\bf 0}$. We start with 
\begin{equation}
\left[ {\widetilde W}_{ij} \right]_{{\bf q} = {\bf 0}} =
\frac{X_i + X_j}{2 \left[ E_i + E_j \right]}
\end{equation}
and for the second derivative, we write
\begin{widetext}
\begin{equation}
\left[
\frac{\partial^2 {\widetilde W}_{ij}}{\partial q_{\ell}^2}
  \right]_{{\bf q} = {\bf 0}} =
\left[
\frac{1}{{\cal D}_{ij}} 
\frac{\partial^2 {\cal N}_{ij}}{\partial q_{\ell}^2}
\right]_{{\bf q} = {\bf 0}}
-
\left[
\frac{2}{{\cal D}_{ij}^2}
\frac{\partial {\cal D}_{ij}}{\partial q_{\ell}}
\frac{\partial {\cal N}_{ij}}{\partial q_{\ell}}
\right]_{{\bf q} = {\bf 0}}
\nonumber
+
\left[
\frac{2 {\cal N}_{ij}}{{\cal D}_{ij}^3}
\left( \frac{\partial {\cal D}_{ij}}{\partial q_{\ell}} \right)^2
\right]_{{\bf q} = {\bf 0}}
-
\left[
\frac{{\cal N}_{ij}}{{\cal D}_{ij}^2}
\frac{\partial^2 {\cal D}_{ij}}{\partial q_{\ell}^2}
\right]_{{\bf q} = {\bf 0}}.
\end{equation}
\end{widetext}
Each one of the four terms in the above expression is evaluated at ${\bf q} = {\bf 0}$ and can
be written in terms of specific expressions that are given below. The numerator is
\begin{equation}
\left[ {\cal N}_{ij} \right]_{{\bf q} = {\bf 0}}
=
X_i + X_j,
\end{equation}
the first derivative of ${\cal N}_{ij}$ is
\begin{equation}
\left[
    \frac{\partial {\cal N}_{ij}}{\partial q_{\ell}}
    \right]_{{\bf q} = {\bf 0}}
=
\frac{Y_j^2}{4T} \frac{\partial E_j}{\partial k_{\ell}}
-
\frac{Y_i^2}{4T} \frac{\partial E_i}{\partial k_{\ell}},
\end{equation}
and the second derivative of ${\cal N}_{ij}$ is
\begin{eqnarray}
\left[
\frac{\partial^2 {\cal N}_{ij}}{\partial q_{\ell}^2}
\right]_{{\bf q} = {\bf 0}}
=
-\frac{X_j Y_j^2}{8T^2} \left( \frac{\partial E_j}{\partial k_{\ell}} \right)^2
+
\frac{Y_i}{8T} \frac{\partial^2 E_i}{\partial k_{\ell}^2}
\nonumber
\\
-
\frac{X_i Y_i^2}{8T^2} \left( \frac{\partial E_i}{\partial k_{\ell}} \right)^2
+
\frac{Y_i^2}{8T} \frac{\partial^2 E_i}{\partial k_{\ell}^2}.
\end{eqnarray}
The denominator ${\cal D}_{ij}$ and its first derivative are
\begin{eqnarray}
\left[ {\cal D}_{ij}
  \right]_{{\bf q} = {\bf 0}}
=
2 ( E_i + E_j),
\\
\left[
\frac{\partial {\cal D}_{ij}}{\partial q_{\ell}}
\right]_{{\bf q} = {\bf 0}}
=
\frac{\partial E_j}{\partial k_{\ell}}
-
\frac{\partial E_i}{\partial k_{\ell}} ,
\end{eqnarray}
while the second derivative of ${\cal D}_{ij}$  is 
\begin{equation}
\left[
\frac{\partial^2 {\cal D}_{ij}}{\partial q_{\ell}^2}
\right]_{{\bf q} = {\bf 0}}
=
\frac{1}{2}
\left[
\frac{\partial^2 E_i}{\partial k_{\ell}^2}
+
\frac{\partial^2 E_j}{\partial k_{\ell}^2} 
\right].
\end{equation}
When the order parameter is zero, that is, $\vert \Delta_0 \vert = 0$,
the energies $E_{1} ({\bf k})$ and $E_{2} ({\bf k})$ become
\begin{eqnarray}
E_1 ({\bf k}) =
\Big\vert \vert \xi_{\bf k}\vert + h_{\bf k} \Big\vert
\\
E_2 ({\bf k}) = 
\Big\vert \vert \xi_{\bf k}\vert  - h_{\bf k} \Big\vert.
\end{eqnarray}
The first derivatives of these energies are 
\begin{eqnarray}
\frac{\partial E_{1} ({\bf k})}{\partial k_{\ell}}
=
S_1 ({\bf k})
\frac{k_{\ell}}{m}
+
\frac{\partial h_{\bf k}}{\partial k_{\ell}},
\\
\frac{\partial E_{2} ({\bf k})}{\partial k_{\ell}}
=
S_2 ({\bf k})
\frac{k_{\ell}}{m}
-
\frac{\partial h_{\bf k}}{\partial k_{\ell}},
\end{eqnarray}
with the functions
$
S_1 ({\bf k})
=
{\rm sgn}\left[ \vert \xi_{\bf k} \vert + h_{\bf k}\right]
{\rm sgn}\left[ \xi_{\bf k} \right]
$
and 
$
S_2 ({\bf k})
=
{\rm sgn}\left[ \vert \xi_{\bf k} \vert - h_{\bf k}\right]
{\rm sgn}\left[ \xi_{\bf k} \right].
$
The derivative of the effective Zeeman field is
\begin{equation}
\frac{\partial h_{\bf k}}{\partial k_{\ell}}
=
\frac{1}{h_{\bf k}}
\frac{\kappa^2}{m^2}
\left(
k_x \delta_{\ell x} + \eta k_y \delta_{\ell y}
\right).
\end{equation}
The second derivatives of the energies are
\begin{eqnarray}
\frac{\partial^2 E_{1} ({\bf k})}{\partial k_{\ell}^2}
=
\frac{S_1 ({\bf k})}{m}
+
\frac{\partial^2 h_{\bf k}}{\partial k_{\ell}^2}
\\
\frac{\partial^2 E_{2} ({\bf k})}{\partial k_{\ell}^2}
=
\frac{S_2 ({\bf k})}{m}
-
\frac{\partial^2 h_{\bf k}}{\partial k_{\ell}^2},
\end{eqnarray}
where the second derivative of the effective field is
\begin{equation}
\frac{\partial^2 h_{\bf k}}{\partial k_{\ell}^2}
=
\frac{1}{h_{\bf k}}
\frac{\kappa^2}{m^2}
\left[
\left( \delta_{\ell x} + \eta \delta_{\ell y} \right)
-
\frac{1}{h_{\bf k}^2}
\frac{\kappa^2}{m^2}
\left(
k_x^2 \delta_{\ell x} + \eta^2 k_y^2 \delta_{\ell y}
\right)
\right].
\end{equation}

Since the diagonal elements ${\widetilde W}_{ii}$ are even functions
of ${\bf q}$ and so are ${\cal N}_{ii}$ and ${\cal D}_{ii}$, their expressions are simpler than in  the
general case discussed above, because the first order derivatives
of ${\cal N}_{ii}$ and ${\cal D}_{ii}$ vanish. The surviving terms involve only
the second derivatives of ${\cal N}_{ii}$ and ${\cal D}_{ii}$ leading to the expression
\begin{equation}
\left[
\frac{\partial^2 {\widetilde W}_{ii}}{\partial q_{\ell}^2}
\right]_{{\bf q} = {\bf 0}} =
\left[
\frac{1}{{\cal D}_{ii}} 
\frac{\partial^2 {\cal N}_{ii}}{\partial q_{\ell}^2}
\right]_{{\bf q} = {\bf 0}}
-
\left[
\frac{{\cal N}_{ii}}{{\cal D}_{ii}^2}
\frac{\partial^2 {\cal D}_{ii}}{\partial q_{\ell}^2}
\right]_{{\bf q} = {\bf 0}}.
\end{equation}
Here, the numerator and denominator functions are 
\begin{equation}
\left[ {\cal N}_{ii} \right]_{{\bf q} = {\bf 0}}
= 2 X_i
\quad
{\rm and}
\quad
\left[ {\cal D}_{ii} \right]_{{\bf q} = {\bf 0}}
= 4 E_i , 
\end{equation}
while their second derivatives are
\begin{eqnarray}
\left[
\frac{\partial^2 {\cal N}_{ii}}{\partial q_{\ell}^2}
\right]_{{\bf q} = {\bf 0}}
& = &
- \frac{X_i Y_i^2}{4T^2}
\left( \frac{\partial E_i}{\partial k_{\ell}} \right)^2
+
\frac{Y_i^2}{4T}
\frac{\partial^2 E_i}{\partial k_{\ell}^2}, 
\\
\left[
\frac{\partial^2 {\cal D}_{ii}}{\partial q_{\ell}^2}
\right]_{{\bf q} = {\bf 0}}
& = &
\frac{\partial^2 E_i}{\partial k_{\ell}^2}.
\end{eqnarray}

The next step in obtaining the $c_{\ell}$ coefficients  is to analyze the functions
${\widetilde \alpha}_{ij}$ and their  second derivatives. We begin by
writing  ${\widetilde \alpha}_{11}$ at
${\bf q} = {\bf 0}$ :
\begin{equation}
\left[
{\widetilde \alpha}_{11}
\right]_{{\bf q} = {\bf 0}}
=
\big\vert u_{\bf k}^2 - \vert v_{\bf k} \vert^2 \big\vert^2
=
\frac{h_z^2}{h_{\bf k}^2}.
\end{equation}
To investigate the second derivative of ${\widetilde \alpha}_{11}$, we write
\begin{equation}
{\widetilde \alpha}_{11} = \gamma_{11} \gamma_{11}^*,
\end{equation}
where the complex function is given by
\begin{equation}
\gamma_{11}
=
u_{{\bf k}_{-}} u_{{\bf k}_{+}} - v_{{\bf k}_{-}} v_{{\bf k}_{+}}.
\end{equation}
In this case, we write the first derivative of ${\widetilde \alpha}_{11}$ as 
\begin{equation}
\frac{\partial {\widetilde \alpha}_{11}}{\partial q_{\ell}}
=
\frac{\partial \gamma_{11}}{\partial q_{\ell}} \gamma_{11}^*
+
\gamma_{11} \frac{\partial \gamma_{11}^*}{\partial q_{\ell}}
\end{equation}
and the second derivative as 
\begin{equation}
\label{eqn:second-derivative-alpha-11}
\frac{\partial^2 {\widetilde \alpha}_{11}}{\partial q_{\ell}^2}
=
\frac{\partial^2 \gamma_{11}}{\partial q_{\ell}^2} \gamma_{11}^*
+
2
\frac{\partial \gamma_{11}}{\partial q_{\ell}} \frac{\partial \gamma_{11}^*}{\partial q_{\ell}}
+
\gamma_{11} \frac{\partial^2 \gamma_{11}^*}{\partial q_{\ell}^2}.
\end{equation}
To explore the symmetry with respect to ${\bf q}$, we express $\gamma_{11}$ in terms of
its odd and even components via the relation $\gamma_{11}
=\gamma_{11,e} + \gamma_{11,o}$, where the even component 
$
\gamma_{11, e} =
\left[ \gamma_{11} ({\bf q}) + \gamma_{11} (-{\bf q}) \right]/2
$
is
\begin{equation}
\gamma_{11, e}
=
u_{{\bf k}_{-}}u_{{\bf k}_{+}} - \vert v_{{\bf k}_{-}} \vert \vert v_{{\bf k}_{+}} \vert
\cos\left( \theta_{{\bf k}_{-}} - \theta_{{\bf k}_{+}} \right)
\end{equation}
and the odd component 
$
\gamma_{11, o} =
\left[ \gamma_{11} ({\bf q}) - \gamma_{11} (-{\bf q}) \right]/2
$
is
\begin{equation}
\gamma_{11, o}
=
i \vert v_{{\bf k}_{+}} \vert \vert v_{{\bf k}_{-}} \vert
\sin\left( \theta_{{\bf k}_{+}} - \theta_{{\bf k}_{-}} \right).
\end{equation}
Expressed via the even $\gamma_{11, e}$ and odd $\gamma_{11,o}$ components,
the second derivative in Eq.~(\ref{eqn:second-derivative-alpha-11}) is 
\begin{equation}
\frac{\partial^2 {\widetilde \alpha}_{11}}{\partial q_{\ell}^2}
=
\frac{\partial^2 \gamma_{11, e}}{\partial q_{\ell}^2} \gamma_{11.e}^*
+
2
\frac{\partial \gamma_{11,o}}{\partial q_{\ell}} \frac{\partial \gamma_{11,o}^*}{\partial q_{\ell}}
+
\gamma_{11,e} \frac{\partial^2 \gamma_{11,e}^*}{\partial q_{\ell}^2}.
\end{equation}
Notice that the even
component is purely real, that is, $\gamma_{11, e}^* = \gamma_{11, e}$, and that
the odd component is purely imaginary, $\gamma_{11,o}^* = -\gamma_{11,o}$.
Use of this property leads to 
\begin{equation}
\frac{\partial^2 {\widetilde \alpha}_{11}}{\partial q_{\ell}^2}
=
2\gamma_{11,e} \frac{\partial^2 \gamma_{11,e}}{\partial q_{\ell}^2}
-
2 \left( \frac{\partial \gamma_{11,o}}{\partial q_{\ell}}\right)^2.
\end{equation}
The contribution from the even term $\gamma_{11,e}$ is
\begin{equation}
\left[
\gamma_{11,e}
\right]_{{\bf q}= {\bf 0} }
=  u_{\bf k}^2 - \vert v_{\bf k} \vert^2
= \frac{h_z}{h_{\bf k}},
\end{equation}
and from its second derivative is 
\begin{equation}
\left[
\frac{\partial^2 \gamma_{11,e}}{\partial q_{\ell}^2}
\right]_{{\bf q} = {\bf 0}}
=
\frac{1}{2}
\left(
\frac{\partial \vert v_{\bf k} \vert}{\partial k_{\ell}}
\right)^2
-
\frac{1}{2} \vert v_{\bf k} \vert \frac{\partial^2 \vert v_{\bf k} \vert}{\partial k_{\ell}^2}
+ \vert v_{\bf k} \vert^2
\left(
\frac{\partial \theta_{\bf k}}{\partial k_{\ell}}
\right)^2,
\end{equation}
while the contribution from the odd term $\gamma_{11, o}$ is
\begin{equation}
\left[
\frac{\partial \gamma_{11,o}}{\partial q_{\ell}}
\right]_{{\bf q} = {\bf 0}}
=
i \vert v_{\bf k} \vert^2
\frac{\partial \theta_{\bf k}}{\partial k_{\ell}}.
\end{equation}

Now, we turn our attention to ${\widetilde \alpha}_{12}$ and its second derivative.
From Eq.~(\ref{eqn:alpha-12}), we notice that $\gamma_{12}$ is explicitly odd in ${\bf q}$
because $\gamma_{12} ({\bf q}) = - \gamma_{12} (-{\bf q})$, since the operation
${\bf q} \to -{\bf q}$ takes ${\bf k}_{-} \to {\bf k}_{+}$ and vice versa, leading to
\begin{equation}
\left[
{\widetilde \alpha}_{12}
\right]_{{\bf q} = {\bf 0}} = 0.
\end{equation}
To calculate the second derivative of ${\widetilde \alpha}_{12}$, we write
\begin{equation}
{\widetilde \alpha}_{12} = \gamma_{12} \gamma_{12}^*,
\end{equation}
where the complex function 
\begin{equation}
\gamma_{12} = u_{{\bf k}_{-}} v_{{\bf k}_{+}} - v_{{\bf k}_{-}} u_{{\bf k}_{+}}.
\end{equation}
We relate $\partial^2 {\widetilde \alpha}_{12}/{\partial q_{\ell}^2}$ to $\gamma_{12}$
and its first and second derivatives via 
\begin{equation}
\frac{\partial^2 {\widetilde \alpha}_{12}}{\partial q_{\ell}^2}
=
\frac{\partial^2 \gamma_{12}}{\partial q_{\ell}^2} \gamma_{12}^*
+
2
\frac{\partial \gamma_{12}}{\partial q_{\ell}} \frac{\partial \gamma_{12}^*}{\partial q_{\ell}}
+
\gamma_{12} \frac{\partial^2 \gamma_{12}^*}{\partial q_{\ell}^2}.
\end{equation}
Given that $\left[ \gamma_{12} \right]_{{\bf q} = {\bf 0}} = 0$ and
$\left[ \gamma_{12}^* \right]_{{\bf q} = {\bf 0}} = 0$, the expression above simplifies to
\begin{equation}
\left[
\frac{\partial^2 {\widetilde \alpha}_{12}}{\partial q_{\ell}^2}
\right]_{{\bf q} = {\bf 0}}
=
2
\left[ \frac{\partial \gamma_{12}}{\partial q_{\ell}} \frac{\partial \gamma_{12}^*}{\partial q_{\ell}} 
  \right]_{{\bf q} = {\bf 0}}
= \left[ \Lambda_{\ell} ({\bf q}) \right]^2,
\end{equation}
where we used the expressions
\begin{equation}
\left[
\frac{\partial \gamma_{12}}{\partial q_{\ell}}
\right]_{{\bf q} = {\bf 0}}
=
e^{i \theta_{\bf k}} \Lambda_{\ell} ({\bf k})
\end{equation}
for the derivatives of $\gamma_{12}$ at ${\bf q} = {\bf 0}$
with the function 
\begin{equation}
\Lambda_{\ell} ({\bf k})
=
u_{\bf k} \frac{\partial \vert v_{\bf k} \vert}{\partial k_{\ell}}
-
\vert v_{{\bf k}} \vert \frac{\partial u_{\bf k}}{\partial k_{\ell}}
+
u_{\bf k} \vert v_{\bf k} \vert
\frac{\partial \theta_{\bf k}}{\partial k_{\ell}}.
\end{equation}

The last information needed is the derivatives of $u_{\bf k}$, $\vert v_{\bf k} \vert$, and
$\theta_{\bf k}$, which are given by
\begin{eqnarray}
\frac{\partial u_{\bf k}}{\partial k_{\ell}}
& = &
-
\frac{1}{2}
\frac{h_z}{h_{\bf k}^3}
\frac{\kappa^2}{m^2}
\frac{( k_x \delta_{\ell x} + \eta k_y \delta_{\ell y})}{(1 + h_z/h_{\bf k})^{1/2}},
\\
\frac{\partial \vert v_{\bf k} \vert}{\partial k_{\ell}}
& = &
\frac{1}{2}
\frac{h_z}{h_{\bf k}^3}
\frac{\kappa^2}{m^2}
\frac{( k_x \delta_{\ell x} + \eta k_y \delta_{\ell y})}{(1 -  h_z/h_{\bf k})^{1/2}},
\\
\frac{\partial  \theta_{\bf k}}{\partial k_{\ell}}
& = &
\eta
\frac{( k_x \delta_{\ell y} - k_y \delta_{\ell x} )}{k_x^2 + \eta^2 k_y^2}.
\label{eqn:uk-vk-thetak-derivatives}
\end{eqnarray}
The long steps discussed above complete the derivation of all the functions needed to compute
the $c_{\ell}$ coefficients for an arbitrary spin-orbit coupling, expressed as a general linear
combination of Rashba and Dresselhaus terms.

As announced earlier, the calculation of  $c_{\ell}$,
defined in Eq.~(\ref{eqn:c-ell-coefficients}),
is indeed very long and requires the use of all the expressions
given from
Eq.~(\ref{eqn:function-f-symmetric}) to
Eq.~(\ref{eqn:uk-vk-thetak-derivatives}).
Despite this complexity, that are a few important comments about the symmetries 
of the $c_{\ell}$ coefficients that are worth mentioning. Given that $c_{\ell}$ determines
the mass anisotropies in the Ginzburg-Landau (GL) theory, we discuss next the anisotropies
of $c_{\ell}$ as a function of the spin-orbit coupling parameters $\kappa$ and $\eta$.
First, in the limit of zero spin-orbit coupling, where $\kappa$
and $\eta$ are equal to zero, all the $c_{\ell}$ coefficients
are identical reflecting the isotropy of the system, that is, $c_x = c_y = c_z$
and reduce to previously known results~\cite{sademelo-1993}.
In this case, the GL effective masses $m_{\ell} = m d_R/c_{\ell}$ are isotropic:
$m_x = m_y = m_z$.
Second, in the limit of $\kappa \ne 0$ and
$\eta = \pm1$, the spin-orbit coupling has the same strength along the $x$ and $y$ directions,
and thus for the Rashba $(\eta = 1)$ or Dresselhaus $(\eta = -1)$ cases,
the coefficients obey the relation $c_x = c_y \ne c_z$.
This leads to effective masses $m_x = m_y \ne m_z$.
Third, in the limit $\kappa \ne 0$,
but $\eta = 0$, corresponding to the ERD case, the coefficients have
the symmetry $c_x \ne c_y = c_z$. Now the effective masses obey the
relation $m_x \ne m_y = m_z$.
Finally, in the case where $\kappa \ne 0$, and $0 \ne  \vert \eta \vert < 1$, all the
$c_{\ell}$ coefficients are different, that is, $c_x \ne c_y \ne c_z$. Therefore, the
effective masses are also different in all three directions: $m_x \ne m_y \ne m_z$.
 
Following an analogous procedure,
we analyze the coefficients $b (q_1, q_2, q_3)$, and $e (q_1, q_2, q_3, q_4, q_5)$ with all 
$q_i = (0,0)$, and define
\begin{equation}
Z_{ij} = X_i + \beta E_i Y_j/2.
\end{equation}
Using the notation $ b (0,0,0) = b (0)$, we obtain 
\begin{widetext}
\begin{equation}
b (0) 
= 
\frac{1}{8V} 
\sum_{\bf k} 
\left[ 
\left(
1 
+ 
\frac{h^4_z}{\xi^2_{\bf k} h^2_{\bf k}}
\right) 
\left( 
\frac{Z_{11}}{E^3_1} 
+ 
\frac{Z_{22}}{E^3_2} 
\right) 
+ 
\frac{2 h^2_z}{\xi_{\bf k} h_{\bf k}} 
\left( 
\frac{Z_{11}}{E^3_1} 
- 
\frac{Z_{22}}{E^3_2} 
\right) 
+ 
\frac{h^4_z}{\xi^3_{\bf k} h^3_{\bf k}} 
\left( \frac{X_1}{E_1} - \frac{X_2}{E_2} 
\right) 
\right],
\end{equation}
which is a measure of the local interaction between two pairing fields.
Using the notation $ f (0,0,0,0,0) = f (0)$, we obtain 
\begin{eqnarray}
f (0) & = & 
\frac{3}{32 V} 
\sum_{\bf k} 
\bigg[ 
-
\left(
1 
+ 
\frac{3 h^4_z}{\xi^2_{\bf k} h^2_{\bf k}} 
\right) 
\left( 
\frac{Z_{11}}{E^5_1} 
+ 
\frac{Z_{22}}{E^5_2} 
\right) 
- 
\frac{h^2_z}{\xi_{\bf k} h_{\bf k}} 
\left(
3 + \frac{h^4_z}{\xi^2_{\bf k} h^2_{\bf k}} 
\right) 
\left( 
\frac{Z_{11}}{E^5_1} - \frac{Z_{22}}{E^5_2} 
\right)   \nonumber   \\
&& \hspace{20mm} 
-
\frac{h^6_z}{\xi^4_{\bf k} h^4_{\bf k}} 
\left( 
\frac{Z_{11}}{E^3_1} + \frac{Z_{22}}{E^3_2} 
\right) 
- 
\frac{h^4_z}{\xi^3_{\bf k} h^3_{\bf k}} 
\left( 
\frac{Z_{11}}{E^3_1} - \frac{Z_{22}}{E^3_2} 
\right)   \nonumber   \\
&& \hspace{20mm} 
+
\frac{\beta^2}{6} 
\left( 
\frac{X_1 Y_1}{E^3_1} + \frac{X_2 Y_2}{E^3_2} 
\right) 
+
\frac{\beta^2 h^2_z}{6 \xi_{\bf k} h_{\bf k}} 
\left( 
\frac{X_1 Y_1}{E^3_1} - \frac{X_2 Y_2}{E^3_2} 
\right) 
- 
\frac{h^6_z}{\xi^5_{\bf k} h^5_{\bf k}} 
\left(
\frac{X_1}{E_1} - \frac{X_2}{E_2} 
\right) 
\bigg],
\end{eqnarray}
which is a measure of the local interaction between three pairing fields.
It is important to mention that in the absence of spin-orbit and 
Zeeman fields, the Ginzburg-Landau coefficients obtained above reduce 
to those reported in the literature~\cite{sademelo-1993}.

\end{widetext}

As we proceed to explicitly write the Ginzburg-Landau action and 
Lagrangian density, we emphasize that in contrast to the standard 
crossover that one observes in the absence of an external 
Zeeman field~\cite{sademelo-1993}, for fixed 
$h_z \neq 0$ it is possible for the system to undergo
a first-order phase transition with increasing 
$1/k_F a_s$. The same applies for fixed $1/k_F a_s$
with increasing $h_z$.  
Thus, while an expansion of 
${\cal S}_F$ to quartic order is sufficient when no Zeeman
fields are present, when 
Zeeman fields are turned on, the fourth-order coefficient $b(0)= b$ may 
become negative. Such a situation requires the analysis of the sixth-order 
coefficient $f (0) =  f$ to describe 
this first-order transition correctly and to stabilize 
the theory since $f > 0$.

The Ginzburg-Landau action in Euclidean space can be written as 
$
{\cal S}_{GL}
= 
\int d t 
\int d^3 {\bf r} 
{\cal L}_{GL} (r),$ \
where $ r \equiv ({\bf r}, t)$. Here,  
the Lagrangian density is
\begin{eqnarray}
{\cal L}_{GL} (r)
&& =  
a 
\vert \Delta (r) \vert^2 
+ 
\frac{b}{2}
\vert \Delta (r)\vert^4  
+ 
\frac{f}{3} 
\vert \Delta (r) \vert^6 \nonumber   \\
& &+  
\sum_{\ell} c_{\ell} \frac{\vert \nabla_\ell \Delta (r) \vert^2 }{2 m} 
- 
i d_0 \Delta^* (r) 
\hspace{.5mm} 
\frac{\partial \Delta (r)}{\partial t},   \
\label{eq:GL}
\end{eqnarray}
where $\ell = \{ x, y, z \}$, $b = b (0)$ and $f = f(0)$. 
A variation of ${\cal S}_{GL}$ with respect to $\Delta^* (r)$ 
via $\delta {\cal S}_{GL} / \delta \Delta^* (r) = 0$ yields
the time-dependent Ginzburg-Landau (TDGL) equation,
\begin{equation}
\left(
-i d_0 \frac{\partial}{\partial t} 
-
\sum_{\ell} c_\ell \frac{\nabla^2_\ell }{2 m} 
+ 
b \vert \Delta \vert^2 
+ 
f \vert \Delta \vert ^4 
+ 
a \right) 
\Delta (r) = 0
\label{eq:sm-GL}
\end{equation}
with cubic and quintic terms, where $\Delta = \Delta (r)$ are dependent on space and time.
This equation describes the spatio-temporal
behavior of the order parameter $\Delta ({\bf r}, t)$ in the
long-wavelength and long-time regime.  

In the static homogeneous case with $b > 0$, Eq.~(\ref{eq:sm-GL}) 
reduces to either the trivial (normal-state) solution 
$\Delta = 0$ when $a > 0$ or to the nontrivial (superfluid state)
$|\Delta| = \sqrt{\vert a\vert /b}$, when $a < 0$. 
The coefficient $d$ provides the timescale of the TDGL equation, 
and thereby determines the lifetime associated with the pairing field 
$\Delta (r)$. This can be seen directly by again considering the homogeneous
case to linear order in $\Delta (r)$, in which case the TDGL equation 
has the solution 
$
\Delta (t) \approx \Delta (0) e^{i a t/d_0}.
$
This last expression can be rewritten more explicitly as 
$
\Delta (t) \approx \Delta (0) e^{-i \omega_0 t} e^{-t/\tau_0},
$
where 
$
\omega_0 
= 
\vert a \vert d_R/ \vert d_0 \vert^2
$
is the oscillation frequency of the pairing field, and  
$
\tau_0 
=
\vert d_0 \vert^2
/
\left(
\vert a \vert d_I
\right)
$
is the lifetime of the pairs, where both $d_R$ and $d_I$ are positive
definite, that is, $d_R > 0$ and $d_I > 0 $.

In the BEC regime, where stable two-body bound states exist, the
imaginary part of $d_0$ vanishes ($d_I = 0$), and the lifetime time
of the pairs is infinitely long. In this case, $d_0 = d_R$ and we 
can define the effective bosonic wave function $\Psi= \sqrt{d_R} \Delta$
to recast Eq.~(\ref{eq:sm-GL}) in the form of the 
Gross-Pitaevskii equation,
\begin{equation}
\left( 
-
i \frac{\partial}{\partial t} 
-
\sum_{\ell} \frac{\nabla^2_\ell }{2 M_{\ell}} 
+ 
U_{2} \vert \Psi \vert ^2 
+
U_{3} \vert \Psi \vert^4
-
\mu_B 
\right)
\Psi (r) 
= 
0,   
\label{eq:sm-GP}
\end{equation}
with cubic and quintic nonlinearities, where $\Psi = \Psi (r)$, to 
describe a dilute Bose gas. Here, $\mu_B = - a / d_R$ 
is the bosonic chemical potential, 
$M_{\ell} = m (d_R / c_{\ell})$ are the anisotropic masses of the bosons, 
and $U_{2} = b / d_R^2$ and $U_{3} = f/d_R^3$ represent contact
interactions of two and three bosons, respectively.
In the Bose regime, the lifetime $\tau$ of the composite boson is  
$\tau \propto 1 / d_I \to \infty$ and the interactions
$U_2$ and $U_3$ are always repulsive, thus leading to a system 
consisting of a dilute gas of stable bosons. In this regime, 
the chemical potential of the bosons is $\mu_B \approx 2 \mu + E_b <  0$,
where $E_b$ is the two-body bound state energy in the presence
of spin-orbit coupling and Zeeman fields obtained from the condition 
$
\Gamma^{-1} ({\bf q}, E - 2\mu)
= 
0
$ 
discussed in the main text. 
Notice that when $\mu_B \to 0^-$, in the absence of boson-boson
interactions, the bosons condense.

\end{document}